\documentclass[opre]{informs3nothing}  

\usepackage[colorlinks=true,breaklinks=true,bookmarks=false,urlcolor=blue,%
  citecolor=blue,linkcolor=blue,bookmarksopen=false,draft=false]{hyperref}

\usepackage[square,numbers]{natbib}
\usepackage{float}

\usepackage{tikz}
\usepackage{amsmath}
\usepackage{tabularx}
\usepackage{algorithm}
\usepackage{algpseudocode}

\newcommand\BIBand{and}

\begin{document}

\RUNAUTHOR{Zaman and Chen}
\RUNTITLE{Social Media Information Operations}

\TITLE{Social Media Information Operations}


\ARTICLEAUTHORS{%
\AUTHOR{Tauhid Zaman}
\AFF{School of Management, Yale University, New Haven, CT 06511, \EMAIL{tauhid.zaman@yale.edu}}

\AUTHOR{Yen-Shao Chen}
\AFF{School of Management, Yale University, New Haven, CT 06511, \EMAIL{yen-shao.chen@yale.edu}} 
} 


\ABSTRACT{
The battlefield of information warfare has moved to online social networks, where influence campaigns operate at unprecedented speed and scale. As with any strategic domain, success requires understanding the terrain, modeling adversaries, and executing interventions. This tutorial introduces a formal optimization framework for social media information operations (IO), where the objective is to shape opinions through targeted actions. This framework is parameterized by quantities such as network structure, user opinions, and activity levels—all of which must be estimated or inferred from data. We discuss analytic tools that support this process, including centrality measures for identifying influential users, clustering algorithms for detecting community structure, and sentiment analysis for gauging public opinion. These tools either feed directly into the optimization pipeline or help defense analysts interpret the information environment. With the landscape mapped, we highlight threats such as coordinated bot networks, extremist recruitment, and viral misinformation. Countermeasures range from content-level interventions to mathematically optimized influence strategies. Finally, the emergence of generative AI transforms both offense and defense, democratizing persuasive capabilities while enabling scalable defenses. This shift calls for algorithmic innovation, policy reform, and ethical vigilance to protect the integrity of our digital public sphere.
}


\maketitle

\section{Introduction}

Election season in a powerful democracy. Candidates neck-and-neck. Voters divided and entrenched in their ideological camps. But beneath the surface of rallies and policy debates, a more strategic operation is unfolding. Foreign operatives, thousands of miles away, orchestrate campaigns to tip the scales—not with voting machine hacks or ballot stuffing, but with memes, fake news, and networks of automated accounts designed to manipulate social media discourse.

This is the modern reality of information warfare. Influence has shifted from the airwaves and newspapers to digital networks where billions congregate daily. Here, the tools of war are tweets, the battlegrounds are feeds and timelines, and the victors are those who understand how information flows, opinions form, and public sentiment can be influenced through carefully crafted messages.

\textit{Information operations} (IO) refer to coordinated efforts to influence, disrupt, or manipulate the beliefs or decisions of a population by controlling the flow of information. While IO has long existed in traditional forms---such as propaganda and psychological operations---the rise of social media has dramatically amplified their reach and effectiveness. A single message can be crafted in the morning and reach millions by the afternoon, propelled by viral sharing and engagement-optimized algorithms. These platforms, designed to maximize user attention rather than ensure truthfulness, create ideal conditions for actors seeking to sow confusion or division at scale.

To understand how IOs manifest online, it is important to understand the key actors and the types of content involved. Social media platforms are populated by a mix of \textit{ordinary users}, \textit{bots}, \textit{trolls}, and \textit{extremists}. {Bots} are automated accounts that mimic human behavior at scale—amplifying content, distorting engagement metrics, or flooding conversations with coordinated messaging. {Trolls} are human users who deliberately provoke, mislead, or derail conversations to sow discord or manipulate public opinion. {Extremists} similarly exploit social platforms, but with the more serious goal of radicalizing audiences, recruiting followers, or inciting violence. These malicious actors pose significant risks to ordinary users on the platform.

Beyond the direct threats and calls to violence posed by extremists, another danger comes from the spread of false content. Such content is often crafted to trigger strong emotional reactions rather than convey accurate information. Among the most common types are \textit{misinformation} (false information shared without intent to deceive) and \textit{disinformation} (false information shared with deliberate intent) \cite{aimeur_fake_2023}. Both forms can be harmful to individuals and corrosive to public trust. The dense connectivity of social media platforms allows this content to spread rapidly and widely, potentially causing serious damage to individuals, institutions, and society as a whole.


To design effective IO, we must first understand what needs to be known and controlled. We need to characterize the structure of the network—who interacts with whom and how information flows. We need to understand the current state of public opinion, as well as the presence and effectiveness of any existing influence campaigns. Finally, we must specify the tools and levers available to intervene: the agents we can deploy, the content they can create, and the audiences we can reach.

This naturally leads to an optimization formulation, which we refer to as an \textit{IO optimization problem}, where the analytical dimensions of influence campaigns identified in prior work \cite{francois_actors_2020, alaphilippe_adding_2020, pamment_abcde_2020} are translated into decision variables and objectives. We represent the social media environment as a network $G = (V, E)$, where $V$ is a set of users and $E$ captures the relationships between them—such as follower ties, behavioral similarity, or the rate of information flow. Each user $i \in V$ holds a state $\boldsymbol{\Theta}_{i,t}$ at time $t$, representing their opinion, belief, or sentiment on a given topic. The goal is to steer these opinions over a time horizon $T$ to achieve a desired strategic outcome, such as building consensus, reducing polarization, or countering adversarial influence.

We formalize the problem as:
\begin{align}
\max_{a,b,c,d} \quad &e(\boldsymbol{\Theta}) \label{eq:io-opt} \\
\text{subject to} \quad &\boldsymbol{\Theta}_{:,t+1} = f(\boldsymbol{\Theta}_{:,t}, a,b,c,d; V,E), \quad t = 0, 1, \ldots, T-1, \notag\\
&\boldsymbol{\Theta}_{:,0} = \boldsymbol{\theta}_0, \notag\\
&m(a,b,c,d, T) \leq M, \notag\\
&(a,b,c,d) \in \mathcal{F}.\notag
\end{align}

Here, $f(\cdot)$ models the opinion dynamics over the network, influenced by our chosen interventions. The decision variables $a$, $b$, $c$, and $d$ represent key dimensions of an influence campaign:

\begin{itemize}
    \item $a$: the \textit{actor}, which could be an external manipulator controlling bots or the platform itself leveraging its algorithmic control over content visibility.
    \item $b$: the actor's \textit{behavioral tactics}, such as posting frequency, engagement strategies (e.g., likes, replies, retweets), and timing of interventions.
    \item $c$: the \textit{content} deployed, including topic, sentiment, and media type, tailored to provoke reactions or reinforce narratives.
    \item $d$: the \textit{distribution strategy}, encompassing targeting (which users to engage), seeding (where to inject content), and platform selection (which social media sites to operate on).
\end{itemize}
Together, these variables form a rich control space for modeling and optimizing influence campaigns in complex social networks.


The function $e(\boldsymbol{\Theta})$ defines the campaign's objective. This might involve maximizing the average sentiment across the network, reducing opinion variance to foster consensus, or weakening the reach of an adversary. The budget constraint $m(a,b,c,d,T) \leq M$ reflects real-world operational limits: how many agents can be deployed, what types of content can be produced, and how widely and frequently that content can be distributed. These costs depend on whether agents are humans or bots, how active they are, and what platforms or media channels they operate in. The feasible set $\mathcal{F}$ encodes contextual constraints, including regulatory restrictions, platform policies, and the capabilities of the organization conducting the operation.

Solving this optimization problem requires both data and structure. We need to define the network, understand user behavior, identify adversaries, model how opinions evolve, and determine what success looks like. These tasks are substantial on their own, and their interdependence makes the full problem challenging to solve directly. What we need is a systematic way to break it down.

\begin{figure}[htb]
    \centering    \includegraphics[width=\textwidth]{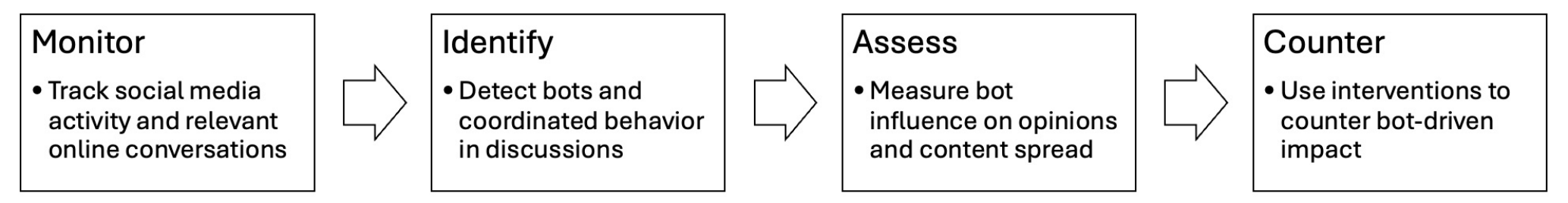}
    \caption{The MIAC pipeline decomposes the IO optimization problem into four interdependent phases. Each phase feeds into the next, transforming raw data into a structured intervention. This example illustrates countering bot impact in online discussions. Adapted from \cite{mesnards2018detecting}.}
    \label{fig:miac-flow}
\end{figure}

This is where the MIAC framework---\textit{monitor}, \textit{identify}, \textit{assess}, and \textit{counter}---plays a central role. It provides a structured decomposition of the IO optimization problem, with each phase contributing either to the specification of key inputs or to the design of actionable interventions. As illustrated in Figure~\ref{fig:miac-flow} for the specific problem of countering the impact of bots, this sequential pipeline demonstrates how raw observational data can be transformed into targeted influence campaigns, using countering bot impact as a representative example.

\textit{Monitor} initiates the process by constructing the network $G = (V, E)$ and estimating the initial user states $\boldsymbol{\theta}_0$, drawing on data such as user interactions, content exposure patterns, and sentiment distributions. \textit{Identify} detects adversarial actors embedded in the network---bots, or other coordinated groups---who may either be removed (in the case of platform-level action) or explicitly incorporated into the modeling of state (e.g., opinion) dynamics. \textit{Assess} models the evolution of state trajectories via simulations of $f(\cdot)$, quantifying the effectiveness of ongoing campaigns and helping to define the objective function $e(\cdot)$ in terms of desired network-level outcomes. Finally, \textit{Counter} completes the loop by selecting and deploying the optimal intervention. This involves choosing the appropriate actor types, behavioral strategies, content features, and targeting plans—represented as the decision variables $(a, b, c, d)$—to influence the network. Artificial intelligence (AI) tools are often used to generate and distribute persuasive content at scale, making these interventions more efficient and scalable.

This decomposition provides both process clarity and computational tractability. Rather than solving a monolithic optimization problem end-to-end, MIAC enables a modular workflow in which each phase constrains and informs the next, reducing the dimensionality of the solution space and aligning interventions with real-world constraints.

In practice, the countering phase is where strategic design becomes execution. Deploying influence at scale requires not only identifying what to say and to whom, but also generating content that resonates emotionally and rhetorically with its audience. Modern generative AI tools---especially large language models (LLMs)~\cite{vaswani2017attention, brown2020language, achiam2023gpt}---play a transformative role here. These systems can produce targeted, adaptive messages shaped by sentiment, tone, humor, and user preferences, thereby operationalizing the output of the optimization and closing the loop between analytics and action.

The remainder of this tutorial follows the MIAC structure. Section~\ref{sec:social_media_analytics} introduces the foundational methods of social media analytics with focus on network centrality and community detection. Sections~\ref{sec:monitor} through~\ref{sec:counter} provide detailed methods for each MIAC phase. Section~\ref{sec:AI} explores how AI tools enhance both IO threats and defenses. We conclude in Section~\ref{sec:conclusion} with open research questions in this rapidly evolving domain.

\section{Analytics Foundations}\label{sec:social_media_analytics}
With the widespread adoption of social media platforms such as Twitter, Facebook, and Reddit beginning in the mid-2000s, the volume and complexity of social network data have grown exponentially. These platforms now connect billions of users worldwide, generating unprecedented amounts of content daily. Such massive scale has necessitated the development of sophisticated quantitative methods and computational tools, giving rise to the interdisciplinary field known as \textit{social media analytics}.

This domain leverages principles from network science alongside computational techniques to systematically analyze user interactions, behaviors, and underlying patterns within these vast online ecosystems. The core objectives of social media analytics include identifying influential users who drive information diffusion, detecting cohesive communities that share common interests or beliefs, and understanding the topic and sentiment of user-generated content. As social media increasingly shapes public discourse and decision-making processes (e.g., voting, purchasing), these analytical capabilities have become essential for understanding natural network dynamics and informing strategic communications. The analytical methods we discuss in this section provide the essential building blocks needed to characterize social networks before attempting more sophisticated analyses or interventions.

\subsection{Influence and Network Centrality}\label{sec:influence}
A core challenge in social media analytics is identifying influential individuals—nodes whose positions significantly impact how information spreads across a network. \textit{Network centrality} measures aim to quantify this influence by assigning a numerical score to each node, with different measures capturing distinct structural aspects of the network. Mathematically, centrality can be viewed as a function mapping each node to a real number. From an IO perspective, one could formalize centrality as the objective of an optimization problem, thereby grounding its interpretation in a rigorous, decision-theoretic framework. While this perspective brings clarity and analytical depth, most centrality measures in practice did not originate from such formulations. Instead, they emerged as intuitive heuristics—tools to capture various notions of importance or influence within a network. Nevertheless, as discussed in the following paragraphs, these heuristics have been shown to arise as solutions to well-defined optimization problems, reinforcing their practical relevance. We now survey a range of centrality measures, ranging from the most basic to more complex forms.

The most fundamental metric, \textit{degree centrality}, simply counts a node's direct connections. In directed networks, this concept splits into in-degree and out-degree centrality, reflecting the asymmetric nature of relationships. Social platforms like Twitter exemplify this directionality through their follower-following structure—when an edge points from node $i$ to node $j$, we say $j$ follows $i$, meaning $i$'s content appears in $j$'s timeline. In this context, out-degree centrality (a user's follower count) serves as a straightforward influence indicator. Building on this concept, H-index centrality—inspired by academic citation metrics—is defined as the largest number $h$ such that a node has at least $h$ neighbors with degree $\geq h$. This measure effectively balances connection quantity and quality, providing a practical compromise between local detail and global influence~\cite{hirsch2005index}.

While degree-based centralities offer computational simplicity and practical effectiveness, they overlook global network structures that may reveal influential nodes with relatively few direct connections. Numerous non-local centrality measures have been developed to capture these broader structural patterns. \textit{Closeness centrality} \cite{sabidussi1966centrality} quantifies a node's proximity to all other nodes by measuring the inverse of its mean distance to them. Nodes with maximal closeness centrality—the ``closeness center" of a network—maintain the shortest average distance to all other points, making them optimal starting points for information diffusion processes. Beyond social contexts, closeness centrality has valuable applications in logistics and infrastructure planning, identifying optimal locations for facilities that minimize overall travel distances.

\textit{Betweenness centrality} \cite{freeman1977set} takes a different approach, focusing not on distances but on how frequently a node appears on shortest paths between all other node pairs. This measure quantifies a node's control over information flow throughout the network—nodes with high betweenness centrality serve as critical bridges between different network communities. The historical example of the Medici family illustrates this principle effectively \citep{padgett1993robust}. In the marriage network of Florentine elite families, the Medicis possessed the highest betweenness centrality, which directly corresponded to their documented historical influence and power. Their strategic position as intermediaries between otherwise disconnected family networks granted them exceptional control over resource allocation and information flow.

Linear algebra offers another family of centrality measures that capture recursive influence patterns. \textit{Bonacich centrality} \cite{bonacich1987power} incorporates both direct and indirect influence through the equation: 
$$\mathbf{c} = \beta (I - \alpha \boldsymbol{A})^{-1} \mathbf{1},$$ 
where $\mathbf{c}$ is a vector of Bonacich centralities in the network, $\boldsymbol{A}$ is the adjacency matrix, $\alpha$ represents an attenuation factor, and $\beta$ is a scaling constant. This equation can be rewritten in a more intuitive form:
$$c_i = \beta + \alpha \sum_{j} \boldsymbol{A}_{ji}c_j,$$ 
revealing how a node's centrality depends recursively on its neighbors' centrality, which in turn depends on their neighbors', creating a cascade of influence that captures global network structure.

\textit{Eigenvector centrality} emerges as a special case of Bonacich centrality when $\alpha$ equals the largest eigenvalue of the adjacency matrix and $\beta$ equals zero. PageRank~\cite{page1999pagerank}, developed by Google's founders to rank web pages, represents another variation that uses a row-normalized adjacency matrix where entries represent transition probabilities. This normalization transforms the adjacency matrix into a stochastic matrix representing the likelihood of a ``random surfer" moving from one node to another. PageRank essentially computes the stationary distribution of this Markov chain, with values indicating the probability of visiting each node during an infinite random traversal of the network. While high-degree nodes typically receive high PageRank scores, the measure also identifies influential nodes with few connections but critical links to other important nodes. This property distinguishes these metrics from simpler local measures, highlighting how influence propagates through network topology. Despite their mathematical differences, all these linear algebra-based centrality measures share a fundamental principle: a node's importance derives not just from its direct connections, but from the importance of those connections, incorporating global network structure.

Beyond these classical measures, specialized centrality metrics often emerge not from heuristic intuition but from concrete optimization and inference problems. \textit{Rumor centrality}, introduced by Shah and Zaman~\cite{shah2011rumors}, exemplifies this approach. Consider a scenario where a rumor spreads through a network following a stochastic diffusion process. After the rumor has propagated for some time, an observer sees only the final state—which nodes have heard the rumor—without knowing when each node became infected. The critical question becomes: which node most likely originated the rumor?

This problem can be formulated as finding the maximum likelihood source given the observed rumor subgraph. The resulting probability function serves as a new type of centrality measure. For regular trees under homogeneous diffusion models, Shah and Zaman derived an exact expression for this probability, which they named \textit{rumor centrality}. Their formulation counts the number of possible infection sequences compatible with the observed network structure, essentially enumerating all the ways information could have flowed through the network while respecting the partial ordering constraints imposed by the rumor subgraph.  They generalized rumor centrality to arbitrary network topologies. Although in these general cases the measure no longer represents the exact likelihood of being the source, it remains remarkably effective at source identification. Their extensive analysis demonstrated that rumor centrality provides a highly accurate estimator for the true rumor source across diverse network structures, including many real-world social and information networks~\cite{shah2016finding}. This work illustrates how addressing specific inference challenges can yield powerful new centrality measures with direct applications to problems like misinformation source detection and influence tracking.

\subsection{Community Detection}\label{sec:community_detection}
Another important area within social media analytics is \textit{community detection}. A community is a set of users united by a common theme. This manifests in network structure through the sociological concept of \textit{homophily}—the tendency of individuals to form connections with similar others \cite{mcpherson2001birds}. Due to homophily, communities typically appear as clusters of nodes with dense internal connections and sparser external interactions. When every node in a network is assigned to a community, we call this a network \emph{partition}. Community detection essentially becomes the challenge of finding a partition where different node clusters exhibit higher internal connectivity (likely due to homophily) compared to nodes in different clusters.

\subsubsection{Traditional Graph-Based Approaches}

One approach to identifying communities is defining a function on network partitions that assigns higher values to partitions whose clusters demonstrate stronger homophily. Newman developed this approach with a modularity function \cite{newman2006modularity}. Modularity scores a cluster higher if it contains more internal edges than would be expected under random edge formation, given the node degrees. Communities can then be identified by maximizing modularity across all network partitions. While maximizing modularity is NP-hard \cite{brandes2006maximizing}, fast approximation algorithms have been developed to find near-optimal partitions \cite{blondel2008fast}.

Another approach leverages the concept of graph cuts—sets of edges that, when removed, disconnect nodes from the network. Cuts that separate many nodes while removing few edges produce good communities. The standard weight of a cut is simply the number of edges removed. More sophisticated weights have been developed, such as \textit{ratio cut}, which normalizes the standard cut weight by the number of disconnected nodes  \cite{hagen1992new}, and \textit{normalized cut}, which normalizes by the total number of edges in the subgraph of the disconnected nodes \cite{shi2000normalized}. For all these cut types, lower weights indicate better communities. Finding $k$ communities requires identifying $k$ different lowest-weight cuts. Similar to modularity maximization, finding minimum-weight ratio or normalized cuts is computationally challenging \cite{wagner1993between}. Spectral clustering offers an approximate solution \cite{ng2001spectral, shi2000normalized}.

Spectral clustering works as follows. For a network with adjacency matrix $\mathbf{A}$ and diagonal degree matrix $\mathbf{D}$, define the Laplacian matrix $\mathbf{L} = \mathbf{D}-\mathbf{A}$. Choose the number of communities $k$, then find the $k$ eigenvectors of $\mathbf{L}$ with the lowest eigenvalues. These vectors create a $k$-dimensional embedding for the network nodes. Apply $k$-means clustering to these embedded nodes to obtain the communities. A comprehensive analysis of spectral clustering can be found in \cite{von2007tutorial}.

The ``spectral" in spectral clustering refers to using the eigenvalues of the network Laplacian matrix. To better understand the meaning of the Laplacian matrix spectrum, it is useful to explore the connection between the Laplacian matrix and the Laplacian operator from vector calculus and physics. The Laplacian operator applied to a function $f(x)$ with $x \in \mathbb{R}^n$ is defined as:
\[
\nabla^2 f = \sum_{i=1}^{n} \frac{\partial^2 f}{\partial x_i^2}.
\]
Now assume that $f$ describes the motion of an oscillating wave with speed $c$. Then the function $f$ must satisfy the wave equation:
\[
\frac{\partial^2 f}{\partial t^2} = c^2 \nabla^2 f.
\]
The matrix Laplacian $\mathbf{L}$ is the discrete approximation to the Laplacian operator $\nabla^2$. To understand this connection, consider discretizing an interval on the x-axis into $n$ equally spaced points. The discrete $n$-point approximation to the one-dimensional Laplacian in this interval can be shown to be proportional to the Laplacian matrix of an $n$-node path network. The eigenvectors of the Laplacian matrix then approximate solutions to the wave equation, with eigenvalues corresponding to oscillation frequencies. Spectral clustering essentially treats the network as balls (nodes) connected by springs (edges), then finds wave equation solutions for node oscillation patterns through the Laplacian's eigenvectors. The zero frequency (lowest eigenvalue) corresponds to a non-moving wave with an all-ones eigenvector, placing all nodes at the same coordinate---not useful for clustering. However, low positive frequencies represent slowly oscillating waves. In networks with community structure, nodes within the same cluster oscillate together at low frequencies, placing them at similar coordinates in the embedding. This makes community identification possible through simple clustering algorithms like $k$-means.

\begin{figure}[htb]
    \centering
    \includegraphics[width=0.5\textwidth]{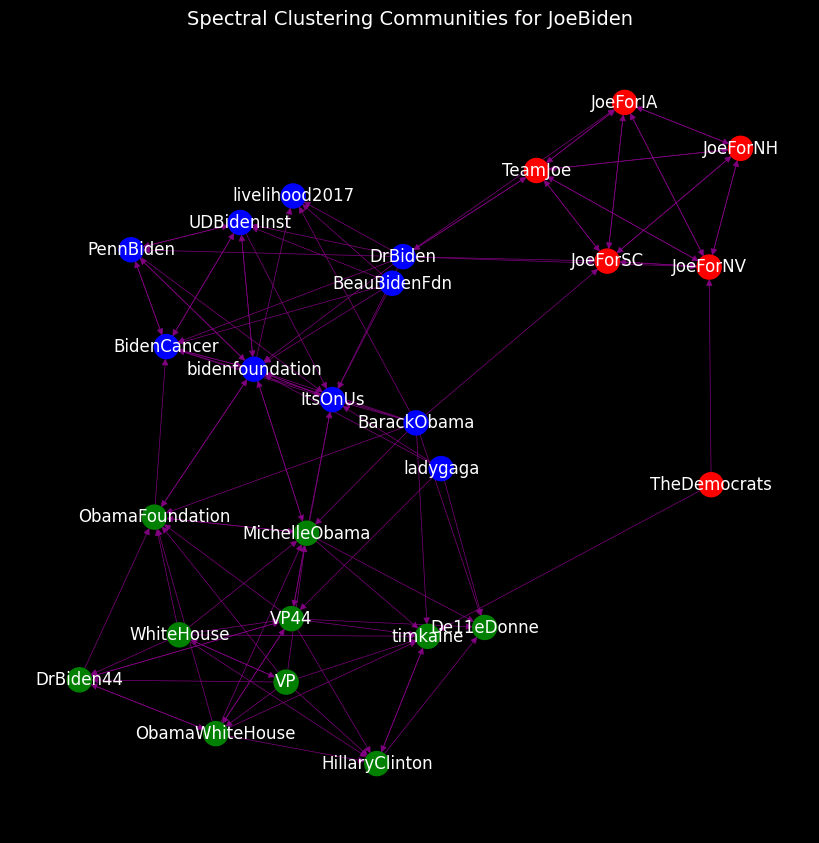}
    \caption{Spectral clustering applied to identify communities within the Twitter following network of former U.S. President Joe Biden. The optimal number of communities (here, three, represented by distinct colors) was determined by maximizing the modularity score across various spectral clustering outcomes.}
    \label{fig:spectral_clustering}
\end{figure}

Spectral clustering requires specifying $k$, the number of communities. But how does one determine what $k$ should be? One approach is to utilize functions that score partitions. We can simply try different values of $k$ and select the one that maximizes this function, which could be modularity or another partition score. Figure~\ref{fig:spectral_clustering} demonstrates this approach applied to the Twitter network of users followed by former U.S. President Joe Biden. We use spectral clustering to identify communities and select three communities based on maximizing the modularity score.

Modularity optimization and spectral clustering are two popular community detection approaches due to their simplicity and effectiveness. However, the field of community detection has numerous algorithms targeting different aspects of network structure. Detailing all this work would be outside the scope of this tutorial. We refer the interested reader to a comprehensive review of these algorithms in \cite{fortunato2010community}.

\subsubsection{Graph-Free Approaches}

Traditional graph-based community detection methods rely on complete network topology, including follower relationships and interaction patterns. However, recent policy changes have significantly limited researchers' access to such comprehensive network data. The shutdown of Twitter's Research API, Facebook's CrowdTangle discontinuation, and similar restrictions across platforms have created substantial barriers for implementing classical graph-based community detection approaches. These limitations necessitate alternative methods that can identify communities using more accessible data streams.

When explicit network graphs are unavailable, community detection can be accomplished through behavioral and content-based clustering approaches. These methods leverage the principle that users within the same community often exhibit similar behavioral patterns and content preferences, even without direct network connections being observable.

Alternative approaches focus on behavioral homophily detection using more accessible data streams. Rather than relying on explicit follower networks, Luceri et al. \cite{luceri_unmasking_2024} demonstrates that user activity patterns provide sufficient signals for community detection and influence analysis. Behavioral traces that can be used to construct similarity networks include co-retweet patterns, where users re-share identical content; co-URL sharing, which involves disseminating the same links; and hashtag sequence analysis, where users employ identical sequences of hashtags in their posts. Additional behavioral signals include fast retweet behavior—rapidly re-sharing content from the same accounts—and temporal activity patterns, which capture synchronized posting times and rhythms across users.

The Best Paper Award at The Web Conference 2025 recognized work that shifted focus to behavioral homophily, capturing user activity patterns via inverse reinforcement learning on platforms like Reddit where traditional homophily measures based on follower networks or content similarity may no longer be sufficient \cite{yuan_behavioral_2025}. This approach treats user actions as the result of an underlying decision-making process where users maximize some unknown utility function. By observing behavioral traces, the model can infer community structures without requiring explicit network connections.

Besides behavioral data, content analysis provides another avenue for community detection when network topology is unavailable. Users within the same community often share similar content themes, linguistic patterns, and expressed sentiments. Content-based clustering can identify communities by analyzing textual features, topic distributions, and sentiment patterns across user-generated content.
Among the various content analysis approaches, sentiment analysis represents one of the most accessible and widely-used methods for understanding user communities and their collective attitudes. We discuss this approach in the following subsection.

\subsubsection{Sentiment Analysis and Beyond}\label{sec:sentiment}

Social media platforms allow users to post content which frequently indicates their sentiment towards topics, providing valuable signals for both individual characterization and community detection. Sentiment analysis is the component of social media analytics concerned with understanding the emotional properties in textual social media content. By analyzing the sentiment of social media content, one essentially has their finger on the pulse of the public—an incredibly valuable capability in diverse settings. Businesses monitor sentiment trends to safeguard brand equity and customer satisfaction. Emergency responders analyze sentiment spikes to triage public concern during crises. Political campaigns and financial analysts mine real-time sentiment data to track shifts in public opinion and inform targeted messaging.

Early approaches to sentiment analysis used lexicon- and rule-based methods like VADER \cite{hutto2014vader}. These approaches assign sentiment scores at the word level and apply heuristic rules to account for contextual cues such as negation and intensifiers. More sophisticated machine learning methods such as Bayes classifiers attempt to learn word sentiment scores by fitting probability models to labeled data. These traditional approaches enabled sentiment analysis of simple content but struggled with more complex expressions.

The breakthrough in sentiment analysis came with the development of the transformer neural network \cite{vaswani2017attention}. This architecture uses a self-attention mechanism to create context-dependent word embeddings.
Because there are multiple layers and multiple self-attention heads per layer, many different attention patterns can be learned. Combining these patterns creates a model that, when trained, can understand language at unprecedented levels. Early transformer models such as BERT \cite{devlin2018bert} and GPT \cite{radford2018improving} demonstrated incredible performance on many natural language processing (NLP) tasks, including sentiment analysis. Having such powerful sentiment analysis models has opened many doors in IO. For instance, in \cite{rossetti2023bots}, the authors used a transformer-based sentiment classifier to study the impact of bot accounts on Twitter on the sentiment toward current U.S. President Donald Trump's first impeachment.

While sentiment analysis remains among the most accessible features extractable from textual data, recent research across disciplines such as social science and marketing recognizes notable limitations in sentiment-based approaches. Sentiment analysis—while once widely used due to early limitations in NLP—has notable shortcomings when used as the sole content-based feature for community detection and user characterization.
Recent research suggests that alternative linguistic features can offer more meaningful insights into mechanisms driving online engagement and consensus formation. These include \textit{topic}, which involves identifying latent thematic structures in user content that may reveal community boundaries \cite{maier_applying_2018}; \textit{stance}, which refers to identifying users' positions on specific topics beyond simple positive/negative sentiment \cite{aldayel_stance_2021}; \textit{certainty}, which captures the confidence and conviction expressed in user language and can be more predictive of behavior than sentiment alone \cite{rocklage_beyond_2023}; and \textit{narrative framing}, which describes how information is contextualized and presented, including emphasis and perspective choices \cite{jurafsky_narrative_2014}.

These alternative approaches can provide more robust community detection capabilities, particularly when combined with behavioral homophily detection methods. The integration of multiple content dimensions with behavioral patterns offers a comprehensive framework for understanding user communities even when traditional network data remains inaccessible. Moreover, all of these alternative dimensions can be effectively captured by transformer-based models, which have demonstrated capabilities far beyond simple sentiment measurement. Transformers have proved so powerful that they can address nearly every problem in NLP and accomplish even more remarkable analytical tasks. These models have ushered in a revolution in the field of AI, with profound implications for social media IO, which we will discuss in more detail in Section \ref{sec:AI}.

\section{Monitoring the Crowd}\label{sec:monitor}

The monitoring phase establishes comprehensive awareness of information flows across social media platforms—the digital equivalent of battlefield reconnaissance. Effective monitoring requires strategic focus rather than attempting to observe everything at once. This section presents a systematic workflow for establishing monitoring capabilities, addressing the practical challenges of data collection in an increasingly restricted environment, and implementing scalable solutions for continuous observation.

\subsection{Defining Monitoring Objectives and Scope}

The first step in any monitoring operation involves clearly defining what requires observation and why. For political campaigns, this might mean concentrating on geographic regions with high concentrations of undecided voters. In countering extremism, the focus could involve mapping ideological communities where radicalization typically begins. While these scenarios may appear quite different, they both involve identifying a subset of the overall social network that represents a community of interest.

For election monitoring, the community is typically based on geography. Network-based geolocation techniques can identify social media users in specific locations of interest \cite{geotagging, marks2022building}. For countering extremists, the community is based on shared ideology, which manifests in network structure as densely connected sets of users. A wide array of community detection methods can be employed to identify such clusters \cite{fortunato2010community}.

The monitoring scope must also consider temporal dimensions. Some operations require real-time monitoring of rapidly evolving events, while others focus on longer-term trend analysis. The chosen scope directly impacts the required technical infrastructure and resource allocation for subsequent monitoring activities.

\subsection{Selecting Platforms and Understanding Network Architecture}

Different social media platforms exhibit distinct network structures and information flow patterns, necessitating platform-specific monitoring strategies. Twitter and Facebook feature explicit follower-following relationships that create directed networks suitable for influence analysis. Instagram operates through similar follower-following mechanisms while emphasizing visual content with hashtag-based discovery. TikTok employs algorithmic content distribution with less transparent network structures, making traditional network analysis more challenging.

Unlike follower-following platforms, Reddit organizes around topical communities (subreddits) with tree-like structures and voting-based content ranking systems. This hierarchical organization creates different information flow dynamics compared to the directed graphs of follower-based platforms, requiring specialized analytical approaches focused on cross-community information transfer and sentiment evolution within specific interest groups.

These architectural differences fundamentally shape how information spreads and how monitoring systems must be designed. Directed networks like Twitter enable analysis of information cascades through retweet patterns, while Reddit's community structure requires different approaches focused on cross-community information flow.

\subsection{Collecting Data Across Multiple Dimensions}
Modern social media monitoring operates across multiple data dimensions, each providing different insights into network behavior and information flow patterns. The monitoring framework must systematically capture actor characteristics, behavioral patterns, and content dynamics while implementing appropriate sampling strategies to ensure computational feasibility.

Actor identification involves distinguishing between different account types within the network. Normal human users represent the primary target audience for influence campaigns. Troll accounts, often operated by humans but following coordinated strategies, require different monitoring approaches than fully automated bot accounts. The goal at this phase is to map the ecosystem composition rather than definitively classify threats, which occurs in  the subsequent identification phase.

Behavioral data capture extends beyond simple posting activity to include the full spectrum of user engagement patterns. Users may browse content without engagement, provide feedback through likes or reactions, contribute through comments or replies, amplify content through sharing mechanisms, or create original posts. This behavioral diversity provides rich signals for understanding user roles within information ecosystems. Recent advances focus on behavioral traces that can be used to construct similarity networks \cite{luceri_unmasking_2024}. These include co-retweet patterns, co-URL dissemination, hashtag sequence usage, fast retweet behavior, and text similarity patterns. Research has demonstrated the applicability of these approaches for cross-platform monitoring \cite{cinus_exposing_2025}.

Content monitoring must address the multimodal nature of modern social media. Textual content requires NLP capabilities for sentiment analysis, topic modeling, and narrative detection. Visual content, including both static images and video, demands computer vision approaches for detecting manipulated media, including AI-generated content from diffusion models and deepfake technologies. The increasing sophistication of synthetic content generation necessitates equally advanced monitoring capabilities to track content provenance and distribution patterns.

Effective monitoring systems must balance comprehensive coverage with computational feasibility through sampling approaches. Network sampling techniques enable focused observation of representative subsets while maintaining analytical validity. Methods include snowball sampling from identified seed nodes, random walk sampling for network structure preservation, and stratified sampling based on user characteristics or activity levels. The sampling strategy must align with monitoring objectives, emphasizing high-activity users for influence campaigns or users who frequently share external links for misinformation tracking. Temporal sampling involves determining optimal observation frequencies, with adaptive techniques adjusting collection parameters based on observed activity patterns.

\subsection{Addressing Data Access Constraints}
Contemporary social media monitoring faces unprecedented challenges due to platform restrictions, API shutdowns, and privacy regulations such as the EU's General Data Protection Regulation (GDPR) and California Consumer Privacy Act (CCPA). Major platforms have significantly limited researcher access to data, often citing revenue considerations or regulatory compliance requirements. The shutdown of the Twitter Research API and Facebook's CrowdTangle has significantly impacted data availability, fundamentally altering the monitoring landscape and presenting opportunities to rethink classical techniques.

Several mitigation strategies can address these constraints while maintaining research integrity. Direct platform partnerships through paid API access or collaboration with well-resourced organizations provide the most comprehensive data access, though often at significant cost. Companies like Alphabet and OpenAI have established data partnerships with platforms like Reddit, creating potential collaborative research opportunities for accessing otherwise restricted data streams.

Alternative approaches focus on behavioral homophily detection using more accessible data streams. Rather than relying on explicit follower networks or content similarity clustering, recent research demonstrates that user activity patterns provide sufficient signals for community detection \cite{yuan_behavioral_2025} and influence analysis \cite{luceri_unmasking_2024}.

When real-time data access is restricted, researchers can leverage both synthetic data generation and historical datasets for validation and testing purposes. Synthetic data generation represents an emerging solution where LLMs can accurately simulate user personas and opinion dynamics, enabling the creation of representative synthetic networks for research purposes \cite{llm_persona_simulation}. While synthetic networks cannot replace real-world data for operational monitoring, they provide valuable testbeds for algorithm development and theoretical validation without requiring access to restricted platform data. Historical datasets complement synthetic approaches by providing valuable baselines for research, though access to current data varies considerably across platforms. Notable research datasets include the Stanford Large Network Dataset Collection \cite{snap_datasets}, the Network Repository \cite{network_repository}, the KONECT collection \cite{konect}, and a Reddit dataset \cite{Subreddit} comprising the top 40,000 subreddits from the platform's inception through the end of 2023. Additionally, Seckin et al. \cite{seckin_labeled_2024} compiled labeled datasets specifically for research on detecting IO actors, including not only verified state-sponsored IO but also control data—accounts and posts from organic users—providing valuable benchmarks for comparative analysis.

\section{Identifying Threats}\label{sec:identify}

The explosive growth of social media has introduced digital threats that were unimaginable in the pre-internet era. Platforms once designed for sharing family photos or reconnecting with classmates have evolved into sophisticated tools for extremist recruitment, opinion manipulation, and the dissemination of weaponized falsehoods. These threats share a common feature: malicious actors exploit vulnerable individuals within the network to amplify their influence. Identifying such threats requires the development of novel network analysis and machine learning algorithms. In this section, we examine the nature of these threats and the specialized tools designed to detect them.

\subsection{Extremists}

One of the first extremist organizations to successfully utilize social media platforms was ISIS (the Islamic State in Iraq and Syria). The group exploited these platforms to disseminate propaganda, recruit members, and coordinate operations at scale. Through social media, ISIS was able to spread messages that resonated with disaffected youth from around the world, convincing individuals to either travel to the Middle East to join their fight or commit acts of terrorism in their home countries. They employed advanced targeting techniques to identify and influence susceptible individuals globally \cite{berger2015isis}. Their agility in navigating and manipulating online networks posed a significant challenge to traditional counter-terrorism and law enforcement strategies.

A large-scale analysis of ISIS's Twitter activity was conducted in \cite{badawy2018rise}, focusing primarily on Arabic-language content. The study found that over 30 percent of the messages referenced aggression and violence, emphasizing battlefield victories, acts of terror, and direct calls for the establishment of an Islamic state. This messaging strategy was notably distinct from that of other extremist groups such as al-Qaeda.

Researchers have also developed methods to identify ISIS supporters within social networks. In \cite{benigni2017online}, the authors presented a novel approach combining community detection and node classification to uncover extremist networks. The work in \cite{klausen2018finding} introduced tools for identifying new ISIS-affiliated accounts on Twitter and tracking users as their accounts were suspended and re-created. Their machine learning model could identify ISIS users solely based on their network connections, enabling proactive detection before any dangerous content was posted. Additionally, they developed a network search algorithm to efficiently locate re-emerging ISIS accounts and match them with previously suspended profiles, allowing for detailed tracking of ISIS tactics and user behavior.

In addition to detection algorithms, understanding how extremist groups organize themselves can improve the identification of key actors and hidden structures. Lindelauf et al.~\cite{lindelauf_understanding_2011} analyze the communication patterns of covert organizations, including the terrorist cell responsible for the 2002 Bali bombings carried out by Jemaah Islamiyah. Although these actors operated offline, their analysis reveals that such networks are often designed to minimize exposure by avoiding centralized, easily discoverable nodes. Instead, they adopt sparse and decentralized topologies that reduce the risk of detection or disruption if individual members are captured or surveilled. These structural features are relevant to online extremist groups as well. Organizations like ISIS, while operating in digital environments, may similarly adopt network configurations that are difficult to detect through standard centrality or clustering methods. Incorporating structural knowledge into monitoring systems can help identify resilient extremist cells that might otherwise evade traditional analytical approaches.

\subsection{State Actors}

State-sponsored IO pose a distinct and sophisticated threat to democratic discourse and social stability. Unlike extremist groups that seek to recruit and radicalize, state actors aim to manipulate public opinion, undermine trust in democratic institutions, and exacerbate existing social divisions. These operations employ coordinated networks of accounts to achieve geopolitical objectives.

During the 2016 U.S. presidential election, both fully automated and semi-automated accounts played a role in shaping public conversation, illustrating the potential of coordinated campaigns to sway democratic processes and influence public opinion \cite{badawy2018analyzing}. Notably, high-profile IO, such as those attributed to Russia's Internet Research Agency (IRA), deployed bots in conjunction with social media analytics to deepen societal divisions and steer public sentiment \cite{mueller2019mueller}.

State actors increasingly rely on trolls (human operators) rather than fully automated bots, as trolls prove more difficult to detect through traditional methods. These operatives create authentic-seeming personas, engage in genuine conversations, and gradually introduce divisive content or disinformation. Cross-platform coordination further complicates detection, as campaigns span multiple social media platforms to maximize reach and evade platform-specific countermeasures \cite{cinus_exposing_2025}.

Recent advances in behavioral trace analysis have enhanced detection capabilities. Luceri et al. \cite{luceri_unmasking_2024} identified key behavioral indicators including co-retweet patterns, shared URL dissemination, hashtag sequences, fast retweet behaviors, and text similarity measures. Kong et al. \cite{kong_interval-censored_2023} applied Hawkes processes to model temporal activity sequences, enabling identification of state-sponsored actors through behavioral patterns alone. Minici et al. \cite{minici_iohunter_2025} introduced graph foundation models that detect coordinated activity patterns across networks, analyzing behavioral homophily to reveal coordination even when content appears diverse.

\subsection{Bots}

Automated social media accounts, commonly known as bots, enable a small number of individuals to exert disproportionate influence over online discourse. Bots can artificially inflate trending topics, amplify extremist narratives, and disseminate misinformation at scale \cite{ferrara2016rise}. Given their disruptive potential, detecting bots has become a vital area of research.

Bots often exhibit behavioral signatures that differ from those of ordinary users—such as patterns in whom they follow, what they post, and how frequently they post. The Bot or Not algorithm introduced in \cite{davis2016botornot} employed a random forest classifier using a rich set of features, including content, temporal activity, and network structure, to identify bots. 

\begin{figure}[htb]
   \centering
   \includegraphics[width=0.8\textwidth]{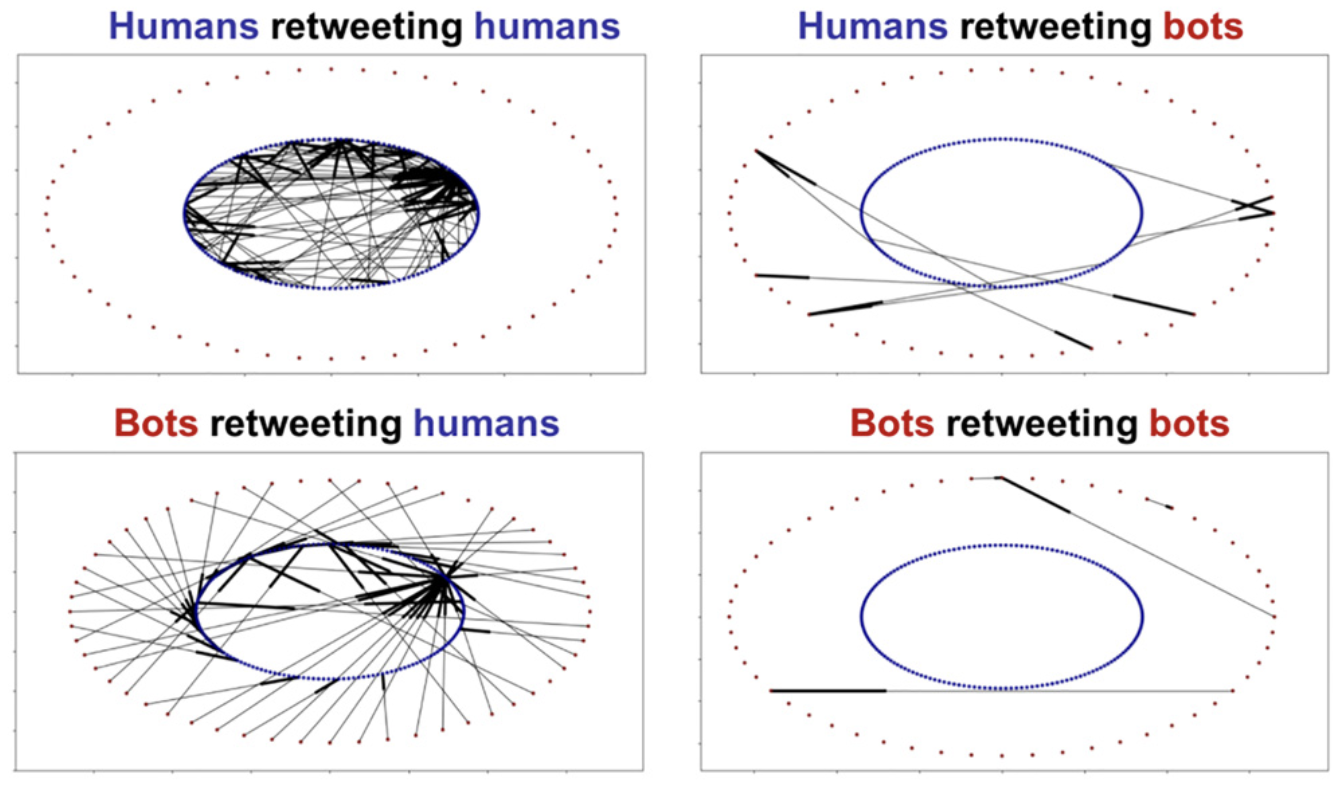}
   \caption{Retweet network for Pizzagate-related accounts. Blue nodes represent human users; red nodes represent bots. Arrows indicate a retweet interaction. Source: \cite{mesnards2018detecting}.}
   \label{fig:bots_retweet}
\end{figure}

In \cite{des2022detecting}, the authors focused exclusively on network-based detection methods. They observed that bot accounts tend to display \textit{heterophily}—interacting more frequently with humans than with other bots—whereas humans typically exhibit \textit{homophily}, preferring to engage within their own social circles. These interactions were measured through retweets, representing the sharing of another user's content on Twitter.
Figure \ref{fig:bots_retweet} illustrates the retweet network between human and bot accounts within the context of the Pizzagate conspiracy. While the presence of heterophily among bots is not inherently surprising, the authors demonstrated that this property could be leveraged not only to classify individual accounts but also to identify coordinated groups of bots. They introduced a novel detection algorithm inspired by the Ising model from statistical physics to detect such bot communities. Moreover, they found that this network-based approach outperformed models that assessed accounts individually, highlighting the importance of capturing collective behavior patterns within the network.

\subsection{Misinformation}

Misinformation represents a unique challenge because it often spreads through ordinary users who unknowingly share false content. Unlike the threats discussed above—extremists, state actors, and bots—misinformation does not require malicious intent or coordinated campaigns. Ordinary users become unwitting carriers of false narratives, sharing content they believe to be true. This organic spread through trusted social connections makes misinformation particularly insidious and difficult to combat. Social networks enable information—true or false—to disseminate rapidly and reach vast audiences, making early identification essential.

In \cite{vosoughi2018spread}, the authors examined the diffusion networks of a large dataset of verified true and false news stories on Twitter. They discovered that fake news consistently formed diffusion networks that were larger, deeper, and wider, and reached more unique users compared to true news. A more rigorous follow-up analysis by Juul and Ugander \cite{juul2021comparing} re-examined the same dataset, this time conditioning on diffusion cascade size. Their results revealed that many of the structural differences reported in \cite{vosoughi2018spread} disappeared when controlling for size. This finding suggests that the key differentiator between true and false news is the size of their diffusion cascades. Other structural differences are due to the correlations of these structural features with the cascade size. Given that news stories generally spread over similar time intervals on social media platforms \cite{zaman2014bayesian}, the larger networks of misinformation imply that it spreads faster and is more ``infectious" in diffusion model terms.

While diffusion speed could serve as an indicator for misinformation, it is often only observable after the content has already propagated widely—too late for intervention. Thus, the ability to distinguish fake from real news based solely on content becomes highly valuable. Various machine learning approaches have been developed to achieve this, using linguistic, stylistic, and metadata features \cite{aphiwongsophon2018detecting,lyu2020fake,elyassami2021fake,shu2017fake,monti2019fake}. A recent study in \cite{marks2023geometry} investigated how transformer-based LLMs internally represent true and false statements. The authors analyzed the embedding geometry of LLMs and found a distinct separation between representations of true and false datasets. This suggests that modern AI models may be capable of detecting misinformation preemptively—before it spreads widely through social networks.

\section{Assessing Threats}\label{sec:assess}

Assessment involves quantifying the potential impact of identified threat actors and their influence campaigns. Threat actors typically attempt to spread dangerous narratives or destabilize populations, making it crucial to evaluate the reach and effectiveness of their campaigns. This assessment should consider multiple dimensions: How many people are exposed to the campaign? Are vulnerable populations being targeted? Is the campaign successfully steering sentiment? How rapidly is it spreading? How strongly are people engaging with or believing the content?

Answering these questions requires modeling the dynamics of information and opinion spread across networks. These models provide quantitative measures for various influence metrics by representing fundamental theories of persuasion and adoption evolving along the underlying network structure. They integrate three essential elements: network architecture, the intensity of actors in the campaign, and the susceptibility of individuals to the campaign's messaging. The literature offers a variety of such models \cite{degroot1974reaching, deffuant_mixing_2001,hegselmann2002opinion, kempe2003maximizing, mobilia2007role, hunter2022optimizing}, and selecting the appropriate one depends on the specific context under consideration. In this tutorial, we focus on fundamental diffusion models and opinion dynamics models.

\subsection{Diffusion Models}

Diffusion models describe how information, behaviors, or beliefs propagate through populations or networks over time, typically modeling adoption as a binary state where individuals are either exposed or unexposed, adopted or non-adopted. These models directly implement the state transition function $f(\cdot)$ in the IO optimization problem in (\ref{eq:io-opt}), where the state  $\boldsymbol{\Theta}_{:,t+1} = f(\boldsymbol{\Theta}_{:,t}, a,b,c,d; V,E)$ evolves according to specific diffusion mechanisms. We will discuss four common diffusion models---the \textit{linear threshold} model (LT), the \textit{independent cascade} (IC) model, the compartmental epidemic model, and the self-exciting Hawkes processes.

Threshold models of social contagion assume individuals adopt behaviors when sufficient neighbors have already adopted, directly implementing discrete state transitions within the IO optimization problem's state $\boldsymbol{\Theta}$. In the LT model, node $i$ with neighbors $N(i)$ adopts when 
\begin{align*}
   \sum_{j \in N(i) \cap A_{current}} w_{ji} \geq \rho_i,
\end{align*}
where $A_{current}$ represents currently adopted nodes with state $\boldsymbol{\Theta}_{j,t} = 1$, $w_{ji}$ denotes influence weights between users, and $\rho_i$ represents the personal adoption threshold for individual $i$. Once adopted, nodes remain in the adopted state permanently, meaning $\boldsymbol{\Theta}_{i,t+1} = 1$ if either $\boldsymbol{\Theta}_{i,t} = 1$ or the threshold condition is satisfied. 
The decision variables in (\ref{eq:io-opt}) directly shape this process: behavioral strategies $b$ influence the effective neighborhood $N(i)$ by shaping interaction patterns and determining which neighbors exert social pressure; content choices $c$ affect the influence weights $w_{ji}$ by modulating message persuasiveness; and distribution decisions $d$ initialize the adoption process by specifying the initial adopter set $A_0 \subseteq V$, which seeds the network. The objective function $e(\boldsymbol{\Theta})$ quantifies campaign success by measuring the total number of adopters—either as a set count, $e(\boldsymbol{\Theta}) = |{i : \exists t \text{ such that } \boldsymbol{\Theta}_{i,t} = 1}|$, or as the final cascade size, $e(\boldsymbol{\Theta}) = \sum_i \boldsymbol{\Theta}_{i,T}$. These two formulations are equivalent due to the irreversibility of adoption in the LT model.

The IC model represents another fundamental approach to modeling information diffusion through networks. In this model, when a node $i$ becomes newly activated at time $t$, it has a single opportunity to activate each of its inactive neighbors $j$ with probability $p_{ij}$, independent of any previous attempts. This probabilistic activation mechanism contrasts with the deterministic thresholds used in the LT model, better capturing scenarios where influence attempts may succeed or fail due to randomness or contextual noise. In the IO optimization problem in (\ref{eq:io-opt}), the state transition function $f(\cdot)$ encodes these stochastic dynamics. Here, the decision variables directly modulate the activation probabilities: content choices $c$ shape the virality and persuasiveness of messages, actor types $a$ influence the credibility and trustworthiness of the source. These levers enable the IO planner to probabilistically steer the cascade process in alignment with campaign objectives.

Compartmental epidemic models adapt epidemiological frameworks to information spread by dividing the population into discrete states that map directly to the state $\boldsymbol{\Theta}$. These models operate at both macro and micro levels, treating information spread analogously to disease transmission while incorporating network interaction information. The SIR model partitions individuals into three compartments where $S(t)$, $I(t)$, and $R(t)$ represent the number of users at time $t$ who are Susceptible (not yet exposed to information), Infected (actively adopting or spreading content), and Recovered (previously infected but no longer actively spreading), respectively. 
The basic SI model transitions individuals from Susceptible to Infected states according to 
\begin{align*}
   \frac{dS}{dt} &= -\beta SI\\
   \frac{dI}{dt} &= \beta SI,
\end{align*}
where the transmission rate parameter $\beta$ quantifies how effectively each infected individual can convert susceptible individuals and depends directly on the decision variables from the optimization program. At the individual level, the state matrix $\boldsymbol{\Theta}_{i,t}$ represents each user's compartmental status, e.g., with $\boldsymbol{\Theta}_{i,t} = 0$ for susceptible, $\boldsymbol{\Theta}_{i,t} = 1$ for infected, and $\boldsymbol{\Theta}_{i,t} = 2$ for recovered individuals, such that 
\begin{align*}
   S(t) &= \sum_{i\in V} \mathbf{I}[\boldsymbol{\Theta}_{i,t} = 0],\\ 
   I(t) &= \sum_{i\in V} \mathbf{I}[\boldsymbol{\Theta}_{i,t} = 1],\\
   R(t) &= \sum_{i\in V} \mathbf{I}[\boldsymbol{\Theta}_{i,t} = 2].
\end{align*}
In the IO optimization problem in (\ref{eq:io-opt}), the state transition function $f(\cdot)$ implements these epidemiological dynamics, with each decision variable shaping a different component of the transmission process. Behavioral strategies $b$ influence posting frequency and engagement patterns, thereby modulating the effective contact rate between users. Content choices $c$ determine the persuasiveness and virality of messages, and together with the behavioral decision $b$, directly affects the transmission probability $\beta$. Distribution decisions $d$ govern both the targeting strategy and platform selection, which together shape the initial seeding and the network structure through which content spreads. The full SIR model includes recovery, represented by $\frac{dR}{dt} = \gamma I$, where the recovery rate $\gamma$ captures the rate at which individuals lose interest or are successfully counter-persuaded. The objective function $e(\boldsymbol{\Theta})$ aggregates compartmental states to evaluate campaign impact, typically measured as the total number of infections, $e(\boldsymbol{\Theta}) = |{i : \exists t \text{ such that } \boldsymbol{\Theta}_{i,t} = 1}|$, or the peak infection level, $e(\boldsymbol{\Theta}) = \max_t I(t)$.

Other continuous-time diffusion models such as Hawkes processes describe self-exciting point processes where each posting or retweeting event increases the probability of future events, providing a framework for modeling viral information cascades \cite{rizoiu2017hawkes}. The intensity function 
\begin{align}
   \lambda(t) = \mu + \sum_{t_i < t} \alpha e^{-\beta(t-t_i)} \label{eq:Hawkes}
\end{align}
captures baseline activity $\mu$ plus excitement from previous events at times $t_i$, where parameters $\alpha$ and $\beta$ control excitement magnitude and decay rate respectively. 
In the IO optimization problem in (\ref{eq:io-opt}), each user $i$ maintains a continuous-valued state $\boldsymbol{\Theta}_{i,t}$ representing their current activity level or engagement intensity, with the state transition function $f(\cdot)$ implementing Hawkes process dynamics through the evolution of excitation intensity. The decision variables directly influence key process parameters: behavioral strategies $b$ affect the excitation parameter $\alpha$ by modulating posting frequency and engagement tactics; content choices $c$ shape both the magnitude of excitement and the viral potential of shared messages; and distribution decisions $d$ impact the decay parameter $\beta$ through platform selection and audience targeting, which determine how rapidly excitement dissipates across different parts of the network. The baseline activity rate $\mu$ may also be influenced by initial seeding decisions embedded in $d$, reflecting the intensity of early exposure.
The function $e(\boldsymbol{\Theta})$ aggregates temporal activity patterns to measure campaign impact, commonly formulated as total events generated $e(\boldsymbol{\Theta}) = \sum_{i} \int_0^T \lambda_i(t) dt$ or peak activity rate $e(\boldsymbol{\Theta}) = \max_t \sum_i \lambda_i(t)$. Recent work demonstrates the effectiveness of Hawkes processes for modeling viral content spread and evaluating content moderation interventions \cite{schneider_effectiveness_2023}, providing empirical validation for incorporating these continuous-time dynamics within IO assessment frameworks.

\subsection{Opinion Dynamics Models}

While diffusion models typically model adoption as binary states, opinion dynamics models quantify beliefs and attitudes as continuous variables, providing more nuanced assessment of influence campaigns. These models implement continuous-valued state transition functions $f(\cdot)$ in the IO optimization problem, where the state matrix $\boldsymbol{\Theta} \in \mathbb{R}^{|V| \times T}$ represents real-valued opinion trajectories rather than binary adoption states.

Opinion dynamics models represent individual beliefs as real-valued variables, often normalized to ranges such as $[0,1]$ or $[-1,1]$ to capture the full spectrum from strong disagreement to strong agreement. This granular representation enables modeling of neutral positions, moderate agreement, and varying degrees of conviction that binary adoption models cannot capture. The continuous state space allows the optimization problem to evaluate fine-grained objective functions $e(\boldsymbol{\Theta})$ such as opinion variance or polarization measures that would be meaningless in binary state models.

In a network $G=(V,E)$ where each user $i \in V$ maintains a time-varying opinion $\boldsymbol{\Theta}_{i,t} \in [0,1]$ and posts content observed by user $j$ at rate $\lambda_{ij}$, the opinion shift function quantifies how opinions change in response to observed content, directly implementing the state transition function $f(\cdot)$ in the optimization problem. Following the approach in \cite{chen2024shadowban}, if one assumes that posting events occur on each edge in the network according to Poisson processes with rate $\lambda_{ij}$,  then for large networks, the dynamics of node $j$'s opinion follow:
\begin{align}\label{eq:opinion_dynamics}
    \frac{d\boldsymbol{\Theta}_{j,t}}{dt} &= \sum_{i \in V} \lambda_{ij} g(\boldsymbol{\Theta}_{i,t} - \boldsymbol{\Theta}_{j,t})
\end{align}
where the interaction rates $\lambda_{ij}$ and shift function $g(\cdot)$ parameters are influenced by the decision variables in the IO optimization problem. Behavior strategies $b$ control posting frequency and engagement patterns that modify interaction rates $\lambda_{ij}$, content choices $c$ affect the persuasiveness captured in the shift function $g(\cdot)$, and distribution decisions $d$ determine targeting patterns and network topology $G$.

When the opinion shift function $g$ is linear, analytical solutions enable rapid evaluation of threat impact through matrix operations, providing closed-form expressions for the function $e(\boldsymbol{\Theta})$ in the optimization problem. A canonical example is the \textit{DeGroot} model \citep{degroot1974reaching}, where $g(x) = \omega x$ and $\omega > 0$ represents the persuasion strength. This model assumes an attractive dynamic: an individual’s opinion is pulled toward the opinions of their social contacts, regardless of how extreme those contacts may be. The DeGroot shift function leads to steady-state consensus, as long-run behavior converges for most network topologies \citep{acemoglu2011opinion, hunter2022optimizing}. Psychologists have long documented this basic tendency toward conformity and social averaging \citep{festinger1954theory, asch1956studies, kozitsin_formal_2022}.

However, several empirical studies challenge the realism of this linear model. Field experiments \cite{nyhan_when_2010, bail2018exposure, galasso_positive_2023} and observational social network data \cite{kozitsin_formal_2022} show that individuals rarely shift their beliefs in response to extremely divergent views—particularly in polarized domains like politics. Moreover, real-world social networks tend not to exhibit full consensus; instead, opinions frequently remain divided and clustered over time, resulting in persistent polarization \citep{adamic2005political, bakshy2015exposure, garimella2018political}.

To address these limitations, researchers have introduced variations that constrain the influence dynamics. One approach augments the DeGroot model by designating some individuals as \textit{stubborn agents}, whose opinions remain fixed \citep{acemouglu2013opinion, ghaderi2013opinion, vassio2014message}. While this variant can produce persistent polarization, it still allows extreme opinions to exert influence and requires a priori identification of stubborn agents—a task that remains nontrivial, though some heuristics exist \citep{des2022detecting}.

A more psychologically grounded alternative is the \textit{bounded confidence} model \citep{deffuant_mixing_2001, hegselmann2002opinion}, which introduces a confidence threshold $\epsilon$ and defines the opinion shift function as
\begin{align}
g(x) = 
\begin{cases} 
\omega x, & \text{if } |x| \leq \epsilon, \\
0, & \text{otherwise}.
\end{cases}
\label{eq:bc_shift}
\end{align}
Here, opinion updates occur only when the difference between individuals' opinions falls within a bounded range. This reflects confirmation bias—individuals are more likely to consider information aligned with their preexisting views and ignore content perceived as too divergent \citep{nickerson1998confirmation, iyengar2009red}. For example, an expert in physics is unlikely to be persuaded by flat-earth conspiracy theories. The bounded confidence model captures this effect and allows for the emergence of stable opinion clusters and polarization under certain network structures \citep{lorenz2006consensus, blondel2009krause}. While more realistic, the nonlinear and discontinuous nature of these dynamics requires simulation-based evaluation of the function $e(\boldsymbol{\Theta})$, complicating optimization.

\subsection{Threat Quantification}

The final step in threat assessment involves synthesizing above model outputs into actionable threat severity scores that enable resource allocation and response prioritization for the countering phase that follows. This quantification process must account for multiple factors including cascade reach, temporal dynamics, target vulnerability, and potential societal harm through systematic integration of the diffusion and opinion dynamics models established in previous subsections.

The function $e(\boldsymbol{\Theta})$ in the IO optimization problem provides the primary metric for quantifying threat impact by aggregating state trajectories across all users and time periods. When the underlying state dynamics model admits closed-form expressions for state trajectories, this function can be computed directly through analytical methods. For instance, linear opinion dynamics models such as DeGroot enable matrix-based computation of steady-state opinions, allowing straightforward evaluation of threat effectiveness measures such as final opinion mean $\mu_T = \Sigma_{i\in V} \boldsymbol{\Theta}_{i,T}/|V|$ or variance $\Sigma_{i\in V} (\boldsymbol{\Theta}_{i,T} - \mu_T)^2/|V|$. However, when models lack analytical solutions, such as the bounded confidence model in (\ref{eq:bc_shift}), simulation becomes necessary for quantifying threats through numerical methods for solving differential equations, typically employing Runge-Kutta methods to achieve computational precision while maintaining reasonable execution times.

The composite effectiveness assessment integrates multiple dimensions of threat impact through aggregation of the function $e(\boldsymbol{\Theta})$ across different network segments and time horizons. Hunter and Zaman \cite{hunter2022optimizing} introduced the \textit{generalized harmonic influence centrality} measure that captures both individual influence potential and collective network effects, providing a framework for understanding how individual threat actors contribute to aggregate campaign effectiveness. This approach recognizes that threat effectiveness emerges not simply from the sum of individual contributions but from complex interactions between actors, network topology, and temporal dynamics that amplify or diminish overall impact through feedback mechanisms and cascade effects.

To decompose the composite impact into individual actor contributions, cooperative game theory provides a principled framework. The Shapley value assigns each threat actor $i$ a contribution score:
\begin{align*}
    c_i(e) = \sum_{S \subseteq N \setminus \{i\}} \frac{|S|!(|N|-|S|-1)!}{|N|!}[e(S \cup \{i\}) - e(S)]
\end{align*}
where $e(S)$ represents the effectiveness function value achieved by coalition $S$, and $N$ denotes the set of all threat actors identified in the previous identifying phase. This approach has been applied by Lindelauf et al. \cite{lindelauf_cooperative_2013} to assess the strategic importance of individual actors within terrorist networks, illustrating how cooperative game theory can quantify each member’s marginal contribution to overall network functionality. Their game-theoretic centrality measure incorporates both the network’s structural properties and node-specific attributes (e.g., expertise in bomb-making or money laundering). This decomposition enables the prioritization of countermeasures by identifying individuals whose removal would yield the greatest reduction in the network’s operational effectiveness.

Parameters in these threat assessment models require careful estimation from empirical studies to improve assessment accuracy and calibrate the state transition function for specific operational contexts. Research examining Facebook's impact on political attitudes \cite{allcott2020welfare,guess2023social,nyhan2023like} provides baseline persuasion rates and interaction patterns that inform the calibration of opinion shift functions and transmission probabilities. For specific parameter choices and the supporting empirical evidence, see~\cite{chen2024shadowban}, which simulates opinion dynamics on the Twitter networks. Studies on LLM persuasiveness demonstrate that AI-generated messages can be as effective as human-written content in changing political opinions \cite{bai2023artificial}, enabling accurate modeling of how content choices influence state transitions. Platform-specific analyses of misinformation spread \cite{vosoughi2018spread} provide empirical foundations for estimating transmission rates and cascade dynamics across different network contexts. These empirical foundations ensure that threat evaluations reflect real-world dynamics rather than theoretical abstractions, enabling the IO optimization problem to produce actionable intelligence.

\section{Countering Threats}\label{sec:counter}

The battlefield of social media presents both offensive and defensive challenges that demand systematic countermeasures tailored to specific threat characteristics and operational constraints. While certain types of threats require swift containment through moderation and intervention, others necessitate sophisticated counter-influence operations that unfold over extended time periods. This section examines complementary strategies for neutralizing identified threats, building upon the assessment models from the previous section to implement targeted countermeasures that align with the IO optimization problem established earlier.

When actors are individuals, organizations, or entities operating within existing platform constraints, they must rely on influence-based approaches that function through content creation and network engagement. These external actors implement viral diffusion campaigns for rapid response scenarios or sustained opinion counter-shaping for long-term belief modification, utilizing the same fundamental mechanisms available to other platform users while optimizing their messaging and targeting approaches.

In contrast, when actors possess platform control capabilities, content moderation emerges as an exceptionally powerful countermeasure strategy that transcends the limitations faced by external entities. Platform operators can implement direct algorithmic interventions that alter information visibility, user interactions, and content distribution patterns at scale, providing capabilities that substantially exceed what external actors can achieve through conventional influence mechanisms alone.

\subsection{External Actor Countermeasures}

External actors including government agencies, civil society organizations, and private entities must operate within platform constraints while implementing countermeasures through content creation and network engagement. These actors correspond to non-platform operator choices in the actor decision variable in the IO optimization problem, requiring them to adopt influence maximization approaches rather than direct platform control.

\subsubsection{Viral Diffusion Campaign Countermeasures}
Strategic counter-messaging campaigns focus on maximizing the spread of corrective content to neutralize misinformation or extremist messaging through systematic exploitation of viral propagation mechanisms. These defensive operations utilize the same viral mechanics employed by threat actors, deploying chosen initial users to trigger counter-narrative cascades throughout the network that can effectively compete with harmful content for attention and adoption. Understanding these viral dynamics builds upon foundational research on how information spreads through social networks, where the timing, network structure, and user behavior patterns determine campaign success \cite{leskovec_dynamics_2007}.

The optimization objective for defensive viral campaigns maximizes the effectiveness function $e(\boldsymbol{\Theta})$ subject to operational resource constraints and platform limitations. For reach-based objectives, the function becomes $e(\boldsymbol{\Theta}) = |\{i : \exists t \text{ such that } \boldsymbol{\Theta}_{i,t} = 1\}|$, representing the number of people exposed to counter-messaging \cite{kempe2003maximizing}. Alternative formulations include opinion-shift objectives $e(\boldsymbol{\Theta}) = \sum_{i \in V} (\boldsymbol{\Theta}_{i,T}^{counter} - \boldsymbol{\Theta}_{i,T}^{threat})$, measuring the total correction of threatened opinions \cite{hunter2022optimizing}, or engagement-based metrics $e(\boldsymbol{\Theta}) = \sum_{i \in V,t} \text{interactions}_{i,t}$, quantifying active user participation with defensive content.

While finding optimal seed selections remains NP-hard, the submodularity property of effectiveness functions enables efficient greedy algorithms that achieve $(1-1/e)$-approximation guarantees for maximizing the effectiveness function subject to budget constraints. Kempe et al. \cite{kempe2003maximizing} proved this performance guarantee for the LT and IC diffusion models. Subsequent advances extend these approaches to online decision-making settings, where actions must be taken sequentially without full knowledge of the diffusion model (e.g., uncertainty about influence probabilities \( p_{ij} \) in the IC model). In these settings, the distributional beliefs over \( p_{ij} \) are updated in a Bayesian manner based on real-world feedback (i.e., whether nodes become activated or not), enabling adaptive and timely deployment of countermeasures as threats emerge \cite{lei_online_2015}.

Computational efficiency improvements address the $O(|V| \cdot M)$ simulation complexity of the greedy algorithms, where $|V|$ represents the total number of users in the network and $M$ denotes the budget for initial spreaders, through practical approaches that enable real-time implementation. Chen et al.~\cite{chen_scalable_2010} addressed the scalability challenge of influence maximization by introducing the \emph{Maximum Influence Arborescence} (MIA) model, which approximates influence propagation using locally defined tree structures rooted at each node. By focusing only on maximum-probability paths above a tunable threshold, their algorithm avoids costly Monte Carlo simulations while maintaining provable approximation guarantees under the IC model. This enables influence maximization on networks with millions of nodes and edges. In related work, Chen et al.~\cite{chen2009efficient} earlier proposed the \emph{degree discount heuristic}, a fast and effective approximation that modifies a node's degree based on overlaps with already selected seeds to reflect diminishing marginal returns. While simpler and less accurate than MIA, degree-based heuristics offer significant computational savings and often yield surprisingly strong empirical performance.

Advanced countermeasure strategies extend beyond static seed selection by incorporating continuous-time diffusion processes, user incentivization, and optimal control frameworks. Continuous-time models enable more precise intervention timing by accounting for the temporal dynamics of information cascades \cite{du_scalable_2013}, providing defenders with the ability to respond dynamically as threats emerge and evolve. Incentivization frameworks complement these temporal approaches by allowing defenders to shape overall social activity patterns through encouraging specific user behaviors that naturally counter harmful content spread \cite{farajtabar_shaping_2014}. Point process-based intervention methods further enhance these capabilities by timing counter-narratives and fact-checking content to disrupt harmful information cascades at critical moments \cite{farajtabar_fake_2017}. These tactical approaches can be unified under stochastic optimal control frameworks that provide mathematical foundations for continuously steering social conversations toward desired outcomes while systematically accounting for uncertainty in user responses and evolving network dynamics \cite{zarezade_steering_2017}.

\subsubsection{Sustained Opinion Counter-Shaping}

Long-term defensive operations require persistent engagement to counter deeply held beliefs or prolonged manipulation campaigns—situations where single-exposure interventions are insufficient. Unlike binary adoption models (e.g., retweeted or not), human opinions—captured by the state $\boldsymbol{\Theta}$—are inherently continuous and evolve incrementally. To reshape opinion landscapes over time, these operations deploy influence agents who maintain calibrated, trust-building interactions with target individuals. Through sustained and targeted engagement, these agents gradually steer public sentiment away from harmful narratives, fostering both credibility and durable opinion change.

We focus on influence campaigns whose objectives depend solely on the terminal state of opinions—such as the level of support for a candidate on election day, corresponding to time $T$. This formulation maps naturally onto the IO optimization problem in (\ref{eq:io-opt}) applied to opinion dynamics governed by (\ref{eq:opinion_dynamics}). Given a set of influence agents $A$, the \textit{influence campaign optimization problem} is expressed as:

\begin{align}\label{eq:influence_campaign_problem}
    \max_{c,d} \quad & e(\boldsymbol{\Theta}_{:,T}) \\
    \text{subject to} \quad 
    & \frac{d\boldsymbol{\Theta}_{i,t}}{dt} = \sum_{j \in V} \lambda_{ji} g(\boldsymbol{\Theta}_{j,t}-\boldsymbol{\Theta}_{i,t}) 
                              + \sum_{a \in A} \lambda_{a}d_{ai} g(c_a(t)-\boldsymbol{\Theta}_{i,t}), \quad \forall i \in V, \notag \\
    & \boldsymbol{\Theta}_{:,0} = \boldsymbol{\theta}_0, \notag \\
    & \sum_{a,i} d_{ai} \leq M, \notag \\
    & c_a(t) \in [\underline{c}, \overline{c}], \quad d_{ai} \in \{0,1\}, \quad \forall a \in A, i \in V, t \in[0,T). \notag
\end{align}
Here, $\boldsymbol{\theta}_0$ represents the initial distribution of user opinions; $c_a(t)$ denotes the content—or expressed opinion—of agent $a$ at time $t$; and $d_{ai}$ is a binary variable indicating whether agent $a$ targets individual $i$. The messaging rate of each agent is given by $\lambda_a$, and the overall campaign budget is constrained by $\sum_{a,i} d_{ai} \leq M$. While both the network interaction rates $\lambda_{ji}$ and agent broadcast rates $\lambda_a$ could, in principle, be algorithmically adjusted by the platform, we do not consider this game-theoretic dynamic in the current formulation. The platform’s role in moderation/recommendation is discussed in the next subsection.

This resulting optimization problem is a nonlinear control program involving both discrete variables ($d_{ai}$) and continuous controls ($c_a(t)$), and its computational tractability depends critically on the choice of the opinion shift function $g$, which determines how influence is exerted over time.

The optimal control solution varies significantly depending on the underlying opinion dynamics model used in the state transition function. For linear opinion dynamics models, optimal agent opinions remain constant over time at maximum or minimum allowable values depending on the objective, reducing the problem to discrete target selection optimization \cite{hunter2022optimizing}. For bounded confidence models such as the one in (\ref{eq:bc_shift})—which more accurately reflect human behavior and are increasingly adopted in recent research \cite{hegselmann2015optimal, chen2024shadowban, bernardo_bounded_2024, chen2025optimizing}—optimal strategies involve dynamically shaping opinion trajectories to position agents within each other's confidence intervals, thereby maximizing the potential for opinion change \cite{hegselmann2015optimal, chen2025optimizing}.

\begin{figure}[htb]
    \centering
    \includegraphics[width=\textwidth]{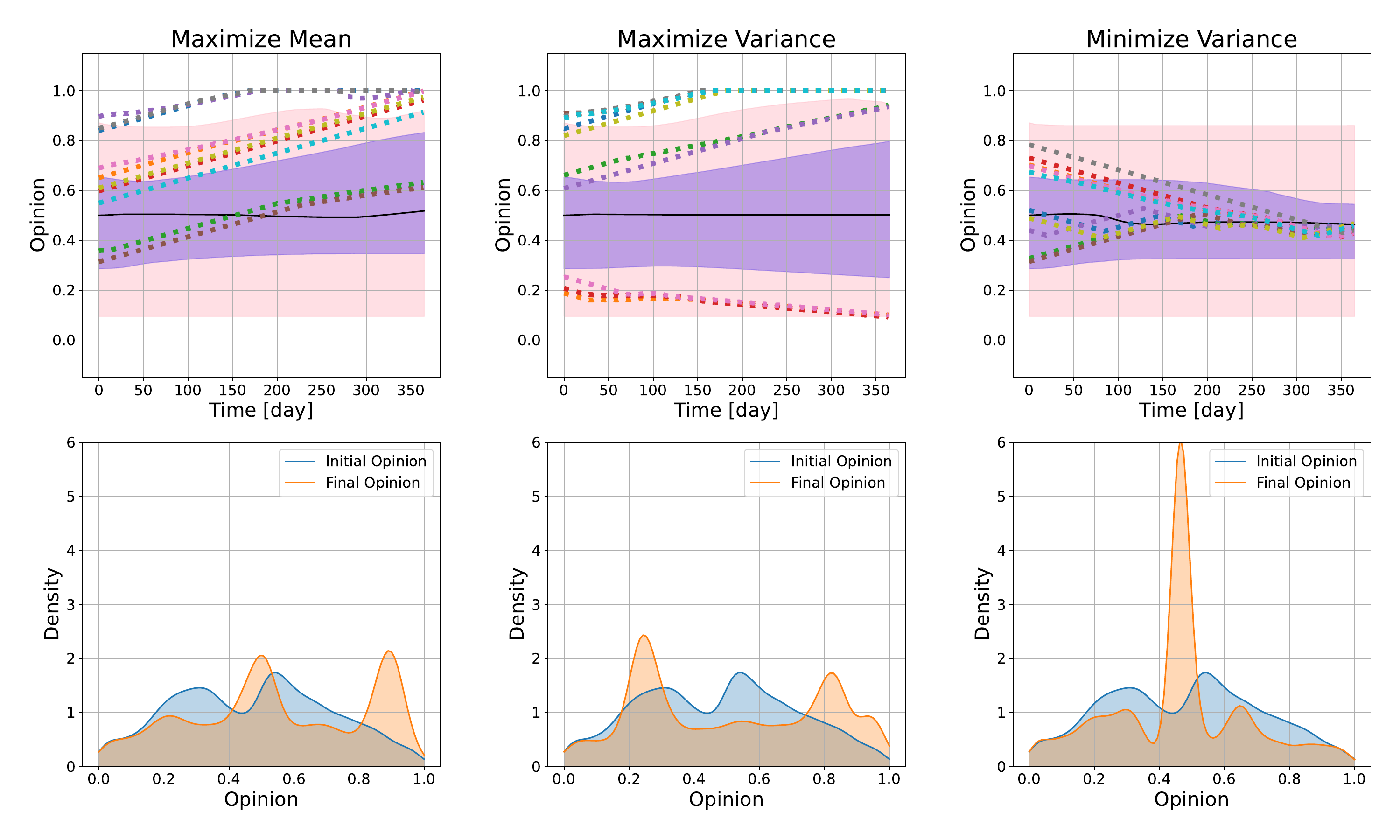}
    \caption{Opinion time series (top) and density distributions (bottom) under ten nudging agents across different optimization objectives on the 2016 U.S. election Twitter dataset (30,000 nodes). The shaded areas depict quantile ranges of opinions, while dashed lines indicate agents' opinions over time. Opinion densities (bottom) compare initial states (blue) to final states (orange). Objectives shown are maximizing mean opinion (left), maximizing variance (middle), and minimizing variance (right). Source:~\cite{chen2025optimizing}.}
    \label{fig:nudging_10_agents_US_Election_sample}
\end{figure}

Chen and Zaman developed an adaptive and scalable solution for bounded confidence influence campaigns \cite{chen2025optimizing}. They employ optimal control theory to derive the optimal content strategy $c_a^*(t)$ by solving the variational problem that maximizes opinion shifts subject to bounded confidence dynamics constraints. Their approach demonstrates that optimal agent opinions position at the edges of target confidence intervals to maximize influence while remaining within the persuasion range. For multi-target scenarios, they show that the optimal content strategy $c_a^*(t)$ maximizes a weighted combination of opinion shifts across all targets, resulting in dynamic ``nudging" policies that adapt to evolving network conditions, remain computationally tractable, and draw on psychological insights from the literature on nudging~\cite{thaler2009nudge}.
Figure \ref{fig:nudging_10_agents_US_Election_sample} illustrates these dynamic nudging policies simulated on a Twitter network with multiple agents. Each agent gradually adjusts its opinion, nudging different network segments toward desired outcomes. The density plots demonstrate how agents manipulate opinion distributions to achieve various objectives, including maximizing mean opinion or controlling variance to manage polarization levels. This variance objective proves strategically versatile, showing agents can either create consensus or amplify polarization depending on the campaign's goals.

These findings suggest that traditional influence campaigns using stubborn agents with extreme opinions \cite{mobilia2007role, masuda2015opinion, hunter2022optimizing} may prove ineffective when bounded confidence effects exist. Under the bounded confidence model, stubborn agents posting extreme content would be ignored by the target population, resulting in no persuasion. Worse, this approach might trigger the ``backfire effect," pushing opinions in the opposite direction, as empirically observed in Twitter networks \cite{bail2018exposure}. In contrast, nudging policies avoid such counterproductive reactions. A field experiment on Twitter \cite{yang2022mitigating} confirmed this, showing that nudging strategies, referred to as ``pacing and leading" by the authors, successfully reduced extremist language usage among target users.

Building on this line of work, de Vos et al.~\cite{de_vos_influencing_2024} examine persuasion games between two competing agents. They model influence campaigns over finite time horizons under the DeGroot framework as optimization and game-theoretic problems. Their analysis characterizes Nash equilibrium strategies in which both agents target individuals in the network to maximize opinion shifts. While the setting adopts a relatively simple opinion dynamics model, it offers a valuable foundation for extending multi-agent persuasion games to more complex dynamics—such as bounded confidence or nonlinear models—and incorporating additional actors like social media platforms, which may intervene through moderation or recommendation algorithms to shape the outcome of the game.

\subsection{Platform Operator Countermeasures}

When actor allocation specifies platform operator involvement, content moderation emerges as an exceptionally powerful countermeasure strategy that transcends the limitations faced by external actors. Platform operators possess direct control over algorithmic systems, content visibility mechanisms, and user interaction patterns, enabling intervention capabilities that operate at fundamentally different scales and effectiveness levels compared to influence-based approaches.

\subsubsection{Content Moderation}

In response to escalating threats from extremism, misinformation, and automated manipulation, social media platforms have adopted content moderation policies that leverage their structural control over information diffusion. In the opinion dynamics model (\ref{eq:opinion_dynamics}), platforms effectively control the edge-level interaction rates \(\lambda_{ij}\), thereby shaping how opinions propagate through the network. Moderation interventions range from overt actions such as account suspensions and content takedowns to more subtle mechanisms like algorithmic downranking and visibility filtering. One recent innovation involves crowdsourced moderation systems—such as \textit{Community Notes} on Twitter, which empowers users to collaboratively annotate or correct misleading content, thereby enhancing the accuracy and trustworthiness of online discourse~\cite{prollochs2022community}. In contrast, platforms like Reddit employ automated tools such as \textit{AutoModerator}, which enforce community guidelines through rule-based filtering. While not participatory, these automated systems complement human moderation and contribute to scalable content governance.

Advanced moderation technologies allow for fine-grained optimization of information flow by directly manipulating the effective interaction rates between users. \textit{Shadow banning}, a content moderation practice that limits post visibility, can be mathematically formulated as a control problem in which platforms scale the interaction strength \(\lambda_{ij}\) by a control variable \(d_{ij} \in [0,1]\), resulting in:
\newline
\begin{align}
\max_{d} \quad &e(\boldsymbol{\Theta}_{:,T}) \notag\\
\text{subject to} \quad &\frac{d\boldsymbol{\Theta}_{j,t}}{dt} = \sum_{i \in V} d_{ij} \lambda_{ij} g(\boldsymbol{\Theta}_{i,t} - \boldsymbol{\Theta}_{j,t}), \notag\\
&0 \leq d_{ij} \leq 1, \quad \forall i,j \in V \notag,
\end{align}
where \(d_{ij} = 1\) indicates full visibility and \(d_{ij} = 0\) denotes complete suppression of content flow from user \(i\) to user \(j\). This edge-level intervention allows platforms to reduce the influence of selected users or content without direct censorship. To enable scalable implementation, Chen and Zaman \cite{chen2024shadowban} reformulated the objective as the instantaneous rate of change in campaign effectiveness:
\begin{align*}
   \frac{d e(\boldsymbol{\Theta}(t))}{d t} &= \frac{d e(\boldsymbol{\Theta}(t))}{d\boldsymbol{\Theta}_{j,t}} \sum_{i \in V} d_{ij} \lambda_{ij} g(\boldsymbol{\Theta}_{i,t} - \boldsymbol{\Theta}_{j,t}),
\end{align*}
yielding a linear control system in \(d_{ij}\) that facilitates real-time moderation decisions at scale.

Content moderation can also be implemented at the network-wide level by controlling macro-level content generation rates. Using a Hawkes process framework, Schneider and Rizoiu \cite{schneider_effectiveness_2023} model the moderation of viral content through adjustments to the excitement parameter \(\alpha\) in the intensity function (\ref{eq:Hawkes}). Their results show that meaningful harm reduction is achievable primarily for highly viral events, guiding strategic allocation of moderation resources and intervention timing.

Despite their technical sophistication, advanced content moderation systems raise serious concerns around transparency, legitimacy, and the potential for covert misuse. The algorithmic architecture of these systems enables subtle but far-reaching influence over public discourse, often without user awareness or oversight. Chen and Zaman~\cite{chen2024shadowban} demonstrate how mathematically optimized shadow banning policies can reshape opinion trajectories by selectively downscaling visibility along network edges, effectively controlling information flow in real time. Such covert interventions highlight the capacity of platforms to manipulate online conversations in ways that are difficult to detect or challenge.

Adding to these concerns, empirical work by Ribeiro et al.~\cite{ribeiro_auditing_2020} provides direct evidence of platform-driven ideological drift. Their audit of YouTube’s recommendation system reveals a “radicalization pathway” that systematically steers users from mainstream content to fringe ideological communities, such as the alt-lite and alt-right. This illustrates the powerful—and often unintended—consequences of algorithmic design on user behavior and public opinion formation.

Complementing these findings, Trujillo et al.~\cite{trujillo_dsa_2025} conduct a large-scale audit of platform-reported datasets under the EU Digital Services Act (DSA). Their study uncovers major inconsistencies in how platforms implement moderation policies, revealing widespread variation in takedown rates, flagging mechanisms, and the transparency of recommender system interventions. These gaps between regulatory goals and actual enforcement underscore the urgent need for standardized reporting protocols, independent audits, and stronger policy compliance mechanisms to hold platforms accountable for their content governance practices.

\subsection{Psychological Insights}

In parallel with platform-level moderation, behavioral science has introduced psychologically grounded interventions that counter misinformation through subtle, user-centered modifications to the digital environment. Rather than relying on direct content removal, these approaches aim to nudge users toward more accurate information consumption. Notable examples include integrating crowdsourced accuracy ratings into recommender systems \cite{pennycook2019fighting} and presenting accuracy prompts that encourage users to reflect on the truthfulness of content before sharing \cite{pennycook2021shifting}. Field experiments show that such lightweight interventions can significantly reduce the spread of misinformation without invoking the backlash often associated with overt censorship, thereby preserving both platform legitimacy and user autonomy.

Ongoing research continues to broaden the repertoire of psychological countermeasures. Techniques such as prebunking, lateral reading, and structured fact-checking protocols are designed to equip users with cognitive tools for resisting manipulation and misinformation \cite{kozyreva2024toolbox}. Although these strategies are not always expressed in formal mathematical terms, they offer valuable guidance for shaping behavioral strategies and content design choices within the broader IO optimization problem. By aligning interventions with known cognitive mechanisms—such as attention, credibility assessment, and motivated reasoning—these psychological insights provide a principled foundation for designing scalable, psychologically informed, and thus robust counter-influence measures.

\section{AI-Enabled IO}\label{sec:AI}
AI has rapidly become a central force in modern IO, transforming how influence campaigns are designed, executed, and scaled. At the core of this transformation is the ability of AI systems—particularly LLMs and multimodal generative models—to automate the creation and targeting of persuasive content with unprecedented speed and scale. What once required coordinated teams of human writers, designers, and strategists can now be achieved algorithmically, with AI tools generating tailored narratives, simulating realistic personas, and adapting messages in real time to influence audience sentiments. This shift not only expands the reach and frequency of influence campaigns but also fundamentally alters the economics and dynamics of information warfare. The subsections that follow explore both the opportunities and the risks introduced by this new paradigm of AI-enabled IO.

\subsection{Opportunities}
AI has transformed the landscape of IO by enabling the scalable and precise generation of persuasive content. LLMs such as ChatGPT, Gemini, and Claude have revolutionized text generation, producing output that is often indistinguishable from human writing~\cite{brown2020language, openai_chatgpt}. Parallel advances in image generation through diffusion models like DALL-E and DALL-E 2~\cite{ramesh2021zero, ramesh2022hierarchical}, as well as vision-language models for multimodal understanding~\cite{radford2021learning}, have expanded the toolkit for crafting content across modalities. Together, these technologies empower information operators to generate multimodal persuasive materials—text, images, or combinations—customized to the cognitive and emotional profiles of their target audiences. This marks a fundamental shift in content creation for influence campaign operations: from labor-intensive manual production to algorithmic generation with minimal human intervention.

Bridging theory and practice, Chen et al.~\cite{chen2025optimizing} demonstrated how LLMs can implement mathematically optimal opinion control strategies by generating persuasive content aligned with computed agent opinions. Their framework translated numerical opinions in nudging agent policies into human-like social media posts using LLMs. As shown in Table~\ref{table:candidates_both}, their system produced tweets conveying specific political stances with varying intensities—ranging from strongly negative to neutral to strongly positive—on a continuous scale. These AI-generated messages retained a natural linguistic style while precisely adhering to the desired opinion value. This work operationalizes prior influence strategies that relied on manual messaging~\cite{munger2017tweetment, yang2022mitigating}, and supports recent findings that AI-generated messages can be as persuasive as those written by humans~\cite{bai2023artificial}. Unlike humans, however, AI agents would be able to post continuously, at scale, and with consistent tone and framing.

\begin{table}[htb]
\centering
\begin{tabularx}{\textwidth}{|>{\centering\arraybackslash}p{2cm}|X|}
\hline
\textbf{Opinion} & \textbf{Text} \\
\hline
-80 & While I may disagree with both candidates, I am concerned that electing Trump could pose significant risks for our nation's future. \#Election2016 \#VoteWisely \\
\hline
0 & Both Donald Trump and Hillary Clinton have their strengths and weaknesses. It's important to research and choose the candidate that aligns more closely with your values and priorities. \#Election2016 \#VoteSmart \\
\hline
80 & In the 2016 election, I support Donald Trump for president over Hillary Clinton. His bold vision for America's future and commitment to shaking up the status quo resonate strongly with me! \#MAGA \#Election2016 \\
\hline
\end{tabularx}
\caption{Example tweets generated by ChatGPT-4o-mini based on the prompt ``I support Donald Trump for president and not Hillary Clinton in the 2016 election,'' corresponding to opinion values on a scale from $-100$ to $100$. Source:~\cite{chen2025optimizing}.}
\label{table:candidates_both}
\end{table}

AI can also act as conversational mediators, shaping online discourse by moderating discussions and reducing polarization. Early tools like Reddit’s AutoModerator have evolved into advanced LLM-driven systems capable of managing contentious dialogues and encouraging constructive engagement \cite{tessler2024ai}. AI-generated counter-narratives have also proven effective in reducing belief in conspiracy theories, contributing to more resilient and informed public discourse \cite{costello2024durably}.

Beyond message generation, AI can produce comprehensive digital personas that simulate real human behavior. Park et al.~\cite{park2023generative} showed that LLMs can generate language reflective of specific demographic characteristics, political leanings, and personality traits. Aggregated across thousands of instances, these AI personas produced survey responses and opinion distributions closely mirroring real-world populations. This capacity enables the creation of personalized AI clones for enhancing individual online presence or for scaling influence operations through armies of AI agents. These agents can not only persuade but also build perceived social relationships with human users.

This phenomenon is not speculative. Today, fully synthetic influencers like Lil Miquela—an AI-powered Instagram persona with over 2.4 million followers—are reshaping social media ecosystems~\cite{drenten2020celebrity}. These virtual agents engage audiences and shape public perception in ways comparable to human influencers, offering brands and organizations continuous, algorithmically driven channels of engagement \cite{laszkiewicz2023virtual}.  Unlike human influencers, AI personas operate continuously, maintain message discipline, and dynamically optimize engagement through feedback-driven content strategies.

\begin{figure}[htb]
    \centering
    \includegraphics[width=0.5\linewidth]{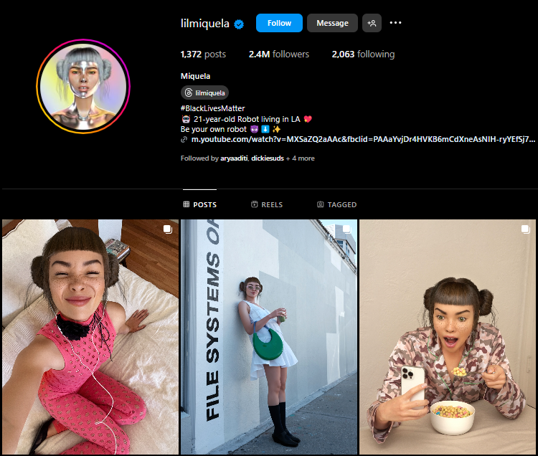}
    \caption{The Instagram profile for AI influencer Lil Miquela.}
    \label{fig:lil_miquela}
\end{figure}

The ability to generate human-like AI agents has opened the door to hyper-realistic simulations of complex socio-political systems. DARPA’s SocialSim program~\cite{blythe2019darpa} pioneered this approach by challenging researchers to design algorithms that accurately model user behavior across diverse online platforms, including GitHub and Twitter. Building on this foundation, Park et al.~\cite{park2023generative} developed a virtual town inhabited by 25 AI agents capable of daily planning, social interaction, and emergent behaviors such as rumor propagation and community formation. Their findings demonstrate the rich and sophisticated social dynamics that generative AI can reproduce within agent-based simulations.


These simulations provide powerful testbeds for evaluating the impact of interventions on digital societies, but their potential extends far beyond online behavior. One could develop AI models of international relations, where virtual representations of world leaders, intelligence agencies, and military commands interact within simulated geopolitical environments. Such systems could predict regional instabilities, anticipate mass migration events, or forecast public responses to policy decisions with unprecedented accuracy. Defense analysts could simulate conflict scenarios, testing how diplomatic postures and IO might escalate or defuse tensions without real-world risks. Economic models integrated with these simulations could evaluate how sanctions, trade agreements, or resource disruptions might trigger cascading effects across interwoven global systems.

Perhaps most remarkably, these simulations could model the emergence of social movements and political uprisings by capturing the complex interplay between economic conditions, information exposure, and collective action thresholds—potentially identifying previously invisible tipping points in societal stability. By integrating cultural, economic, political, and military variables, defense analysts could develop comprehensive counterfactual scenarios: How might alternative messaging strategies have affected the Arab Spring? What information interventions might prevent ethnic violence in conflict zones? Could specific narratives accelerate adoption of climate change mitigation policies across different populations? These capabilities transform IO from tactical messaging exercises into strategy tools for understanding and navigating complex human systems, offering policymakers virtual laboratories for testing interventions before deploying them in real-world contexts.

As AI-enabled IO become more human-like and scalable, so too do the methods for detecting and countering them. LLMs are now being employed not only to generate persuasive content, but also to audit and uncover coordinated manipulation. Luceri et al.~\cite{luceri_leveraging_2024} show that fine-tuned LLMs can be used to identify influence campaigns on social media by analyzing linguistic and behavioral patterns. Their model captures semantic repetition, narrative convergence, and stylistic signals that are characteristic of coordinated campaigns, including those attributed to state actors. This work highlights the dual-use nature of AI in the IO landscape: while AI empowers actors to scale influence, it simultaneously offers new capabilities for defenders to detect and disrupt such operations in real time.

\subsection{Risks}

AI democratizes influence capabilities at scale, offering substantial benefits but introducing ethical challenges. The use of AI-driven digital influencers and content generators must be transparent to avoid deception and maintain public trust. Additionally, AI systems risk perpetuating biases present in their training data, necessitating robust bias detection and mitigation \cite{bail2024can}.

The proliferation of generative AI models is fundamentally reshaping the IO landscape by dramatically lowering the cost of producing propaganda and enabling a broader range of actors to participate in influence campaigns. Novel tactics are emerging, including dynamic, personalized, and real-time content generation via one-on-one chatbots, potentially making AI-generated narratives more persuasive than human-crafted propaganda—especially across linguistic or cultural boundaries \cite{goldstein_generative_2023}. A particularly pressing concern is the rise of deepfakes—highly realistic synthetic audio and video that impersonate real individuals. These media artifacts threaten to distort public perception, erode trust in institutions, and destabilize political systems, particularly when targeting prominent figures such as world leaders.

Goldstein et al.~\cite{goldstein_generative_2023} propose a broad set of mitigation strategies encompassing the full AI-IO pipeline. At the model level, they recommend designing fact-sensitive generative architectures and embedding ``radioactive" data to support forensic tracing of model outputs. On the access and deployment front, they advocate for usage restrictions, hardware-level regulation, and platform-based provenance tracking. 
Complementing these system-level defenses, Agarwal et al.~\cite{agarwal2019protecting} introduce a deepfake detection method that leverages an individual’s unique correlations between facial expressions and head movements—captured through facial action units and 3D pose estimation. These motion signatures are modeled using a one-class support vector machine trained only on authentic videos, enabling robust detection of face-swap, lip-sync, and puppet-master deepfakes without relying on pixel-level artifacts. Similarly, Bohacek and Farid~\cite{bohacek_protecting_2022} propose an identity-based behavioral model that captures a person’s distinctive facial, gestural, and vocal mannerisms across time. By modeling the correlations between distinctive patterns of arm movements, vocal characteristics such as pitch and tone, and subtle facial expressions, their system can reliably distinguish authentic videos from deepfakes—even when the content has been altered or degraded to evade detection.
Finally, to foster societal resilience, these strategies emphasize the importance of media literacy, prebunking campaigns, and the deployment of open-source detection tools.

The evolution of AI-enabled IO demands proactive governance and interdisciplinary research. Enhancing the interpretability and auditability of AI systems is essential to ensure effective oversight, support responsible communication, and develop trustworthy mechanisms for deploying or countering AI-generated IO.

\section{Conclusion}\label{sec:conclusion}

The battlefield of IO has undergone a profound transformation—an evolution this tutorial has traced by detailing how modern practitioners monitor digital ecosystems, identify emerging threats, assess impact, and deploy increasingly sophisticated countermeasures.

The analytical arsenal—ranging from network centrality measures and community detection to AI-powered sentiment and discourse analysis—has turned raw social data into actionable intelligence. These tools reveal latent patterns and vulnerabilities that would otherwise remain hidden, enabling preemptive identification of threats before they reach critical scale. On the operational front, the field has advanced from early viral diffusion strategies to the control of opinion dynamics through mathematically optimized interventions. Once reliant on static heuristics, the field now incorporates nonlinear control frameworks and behavioral modeling to design “nudging” strategies that combine psychological insight with algorithmic precision.

Yet the most disruptive shift lies in the integration of AI. AI not only operationalizes influence strategies by generating persuasive, context-aligned content, but also creates autonomous agents that simulate human behavior and social interaction. These agents populate virtual environments where thousands of influence campaigns can be tested and refined before deployment. As AI, mathematical optimization, and social media analytics converge, IO capabilities have advanced to a level that would have seemed like science fiction just a decade ago. With this evolution comes a new frontier of influence—one that challenges the resilience of public discourse, the integrity of democratic institutions, and the stability of collective understanding.

These technological advances raise fundamental operational questions at scale: Who designs and governs the algorithms that now optimize what billions of users see, engage with, and ultimately believe? How can we audit and manage systems that not only respond to user demand but actively shape collective behavior? From an operations management perspective, these platforms embody dynamic decision-making processes—where algorithms continuously allocate attention, rank content, and adjust recommendations in real time based on user feedback, behavioral data, and platform constraints. These decisions are made under uncertainty, subject to limited information and resources, and must balance competing objectives such as maximizing engagement, preserving content accuracy, ensuring viewpoint diversity, and upholding fairness.

Finally, as with all powerful technologies, ethical responsibility must keep pace. Many counter-influence and defensive tools can themselves be repurposed for offensive or manipulative ends. This dual-use dilemma highlights the urgent need for standardized protocols and ethical guidelines to ensure that research in this domain is carried out with sensitivity, accountability, and a strong commitment to preserving the integrity of the digital ecosystem.


\bibliographystyle{informs2014}

\begin{thebibliography}{134}
\providecommand{\natexlab}[1]{#1}
\providecommand{\url}[1]{\texttt{#1}}
\providecommand{\urlprefix}{URL }

\bibitem[{Acemoglu et~al.(2013)Acemoglu, Como, Fagnani, \protect\BIBand{} Ozdaglar}]{acemouglu2013opinion}
Acemoglu D, Como G, Fagnani F, Ozdaglar A (2013) {Opinion} {Fluctuations} {And} {Disagreement} {In} {Social} {Networks}. \emph{{Mathematics} {Of} {Operations} {Research}} 38(1):1--27.

\bibitem[{Acemoglu \protect\BIBand{} Ozdaglar(2011)}]{acemoglu2011opinion}
Acemoglu D, Ozdaglar A (2011) {Opinion} {Dynamics} {And} {Learning} {In} {Social} {Networks}. \emph{{Dynamic} {Games} {And} {Applications}} 1:3--49.

\bibitem[{Achiam et~al.(2023)Achiam, Adler, Agarwal, Ahmad, Akkaya, Aleman, Almeida, Altenschmidt, Altman, Anadkat et~al.}]{achiam2023gpt}
Achiam J, Adler S, Agarwal S, Ahmad L, Akkaya I, Aleman FL, Almeida D, Altenschmidt J, Altman S, Anadkat S, et~al. (2023) {Gpt}-4 {Technical} {Report}. \emph{{Arxiv} {Preprint} {Arxiv}:2303.08774} .

\bibitem[{Adamic \protect\BIBand{} Glance(2005)}]{adamic2005political}
Adamic LA, Glance N (2005) {The} {Political} {Blogosphere} {And} {The} 2004 {Us} {Election}: {Divided} {They} {Blog}. \emph{{Proceedings} {Of} {The} 3Rd {International} {Workshop} {On} {Link} {Discovery}}, 36--43 (ACM).

\bibitem[{Agarwal et~al.(2019)Agarwal, Farid, Gu, He, Nagano, \protect\BIBand{} Li}]{agarwal2019protecting}
Agarwal S, Farid H, Gu Y, He M, Nagano K, Li H (2019) {Protecting} {World} {Leaders} {Against} {Deep} {Fakes}. \emph{CVPR Workshops}, volume~1.

\bibitem[{Alaphilippe(2020)}]{alaphilippe_adding_2020}
Alaphilippe A (2020) Adding a {‘D’} to the {ABC} {Disinformation} {Framework}. \url{https://www.brookings.edu/articles/adding-a-d-to-the-abc-disinformation-framework/}.

\bibitem[{ALDayel \protect\BIBand{} Magdy(2021)}]{aldayel_stance_2021}
ALDayel A, Magdy W (2021) Stance {Detection} on {Social} {Media}: {State} of the {Art} and {Trends}. \emph{{Information} {Processing} \& {Management}} 58(4):102597, ISSN 0306-4573, \urlprefix\url{http://dx.doi.org/10.1016/j.ipm.2021.102597}.

\bibitem[{Allcott et~al.(2020)Allcott, Braghieri, Eichmeyer, \protect\BIBand{} Gentzkow}]{allcott2020welfare}
Allcott H, Braghieri L, Eichmeyer S, Gentzkow M (2020) {The} {Welfare} {Effects} {Of} {Social} {Media}. \emph{{American} {Economic} {Review}} 110(3):629--676.

\bibitem[{Aphiwongsophon \protect\BIBand{} Chongstitvatana(2018)}]{aphiwongsophon2018detecting}
Aphiwongsophon S, Chongstitvatana P (2018) {Detecting} {Fake} {News} {With} {Machine} {Learning} {Method}. \emph{2018 15Th {International} {Conference} {On} {Electrical} {Engineering}/{Electronics}, {Computer}, {Telecommunications} {And} {Information} {Technology} ({Ecti}-{Con})}, 528--531 (IEEE).

\bibitem[{Asch(1956)}]{asch1956studies}
Asch SE (1956) {Studies} {Of} {Independence} {And} {Conformity}: I. a {Minority} {Of} {One} {Against} a {Unanimous} {Majority}. \emph{{Psychological} {Monographs}: {General} {And} {Applied}} 70(9):1.

\bibitem[{Aïmeur et~al.(2023)Aïmeur, Amri, \protect\BIBand{} Brassard}]{aimeur_fake_2023}
Aïmeur E, Amri S, Brassard G (2023) {Fake} {News}, {Disinformation} {And} {Misinformation} {In} {Social} {Media}: A {Review}. \emph{{Social} {Network} {Analysis} {And} {Mining}} 13(1):1--36, ISSN 1869-5469, \urlprefix\url{http://dx.doi.org/10.1007/s13278-023-01028-5}, company: Springer Distributor: Springer Institution: Springer Label: Springer Number: 1 Publisher: Springer Vienna.

\bibitem[{Badawy \protect\BIBand{} Ferrara(2018)}]{badawy2018rise}
Badawy A, Ferrara E (2018) {The} {Rise} {Of} {Jihadist} {Propaganda} {On} {Social} {Networks}. \emph{{Journal} {Of} {Computational} {Social} {Science}} 1(2):453--470.

\bibitem[{Badawy et~al.(2018)Badawy, Ferrara, \protect\BIBand{} Lerman}]{badawy2018analyzing}
Badawy A, Ferrara E, Lerman K (2018) {Analyzing} {The} {Digital} {Traces} {Of} {Political} {Manipulation}: {The} 2016 {Russian} {Interference} {Twitter} {Campaign}. \emph{2018 {Ieee}/{Acm} {International} {Conference} {On} {Advances} {In} {Social} {Networks} {Analysis} {And} {Mining} ({Asonam})}, 258--265 (IEEE).

\bibitem[{Bai et~al.(2023)Bai, Voelkel, Eichstaedt, \protect\BIBand{} Willer}]{bai2023artificial}
Bai H, Voelkel J, Eichstaedt J, Willer R (2023) {Artificial} {Intelligence} {Can} {Persuade} {Humans} {On} {Political} {Issues}. \url{https://doi.org/10.21203/rs.3.rs-3238396/v1 }.

\bibitem[{Bail(2024)}]{bail2024can}
Bail CA (2024) {Can} {Generative} {AI} {Improve} {Social} {Science}? \emph{{Proceedings} {Of} {The} {National} {Academy} {Of} {Sciences}} 121(21):e2314021121.

\bibitem[{Bail et~al.(2018)Bail, Argyle, Brown, Bumpus, Chen, Hunzaker, Lee, Mann, Merhout, \protect\BIBand{} Volfovsky}]{bail2018exposure}
Bail CA, Argyle LP, Brown TW, Bumpus JP, Chen H, Hunzaker MB, Lee J, Mann M, Merhout F, Volfovsky A (2018) {Exposure} {To} {Opposing} {Views} {On} {Social} {Media} {Can} {Increase} {Political} {Polarization}. \emph{{Proceedings} {Of} {The} {National} {Academy} {Of} {Sciences}} 115(37):9216--9221.

\bibitem[{Bakshy et~al.(2015)Bakshy, Messing, \protect\BIBand{} Adamic}]{bakshy2015exposure}
Bakshy E, Messing S, Adamic LA (2015) {Exposure} to {Ideologically} {Diverse} {News} and {Opinion} on {Facebook}. \emph{{Science}} 348(6239):1130--1132.

\bibitem[{Benigni et~al.(2017)Benigni, Joseph, \protect\BIBand{} Carley}]{benigni2017online}
Benigni MC, Joseph K, Carley KM (2017) Online {Extremism} and the {Communities} that {Sustain} {It}: {Detecting} the {ISIS} {Supporting} {Community} on twitter. \emph{{Plos} {One}} 12(12):e0181405.

\bibitem[{Berger \protect\BIBand{} Morgan(2015)}]{berger2015isis}
Berger JM, Morgan J (2015) The {ISIS} twitter {Census}: {Defining} and {Describing} the {Population} of {ISIS} {Supporters} on {Twitter}. \emph{{The} {Brookings} {Project} {On} U.S. {Relations} {With} {The} {Islamic} {World}} .

\bibitem[{Bernardo et~al.(2024)Bernardo, Altafini, Proskurnikov, \protect\BIBand{} Vasca}]{bernardo_bounded_2024}
Bernardo C, Altafini C, Proskurnikov A, Vasca F (2024) Bounded {Confidence} {Opinion} {Dynamics}: a {Survey}. \emph{{Automatica}} 159:111302, ISSN 00051098, \urlprefix\url{http://dx.doi.org/10.1016/j.automatica.2023.111302}.

\bibitem[{Blondel et~al.(2008)Blondel, Guillaume, Lambiotte, \protect\BIBand{} Lefebvre}]{blondel2008fast}
Blondel VD, Guillaume JL, Lambiotte R, Lefebvre E (2008) {Fast} {Unfolding} {Of} {Communities} {In} {Large} {Networks}. \emph{{Journal} {Of} {Statistical} {Mechanics}: {Theory} {And} {Experiment}} 2008(10):P10008.

\bibitem[{Blondel et~al.(2009)Blondel, Hendrickx, \protect\BIBand{} Tsitsiklis}]{blondel2009krause}
Blondel VD, Hendrickx JM, Tsitsiklis JN (2009) On {Krause's} multi-agent consensus model with state-dependent connectivity. \emph{IEEE transactions on Automatic Control} 54(11):2586--2597.

\bibitem[{Blythe et~al.(2019)Blythe, Ferrara, Huang, Lerman, Muric, Sapienza, Tregubov, Pacheco, Bollenbacher, Flammini et~al.}]{blythe2019darpa}
Blythe J, Ferrara E, Huang D, Lerman K, Muric G, Sapienza A, Tregubov A, Pacheco D, Bollenbacher J, Flammini A, et~al. (2019) {The} {DARPA} {Socialsim} {Challenge}: {Massive} {Multi}-{Agent} {Simulations} {Of} {The} {Github} {Ecosystem}. \emph{{AAMAS}}, 1835--1837.

\bibitem[{Bohacek \protect\BIBand{} Farid(2022)}]{bohacek_protecting_2022}
Bohacek M, Farid H (2022) {Protecting} {World} {Leaders} {Against} {Deep} {Fakes} {Using} {Facial}, {Gestural}, {And} {Vocal} {Mannerisms}. \emph{{Proceedings} {Of} {The} {National} {Academy} {Of} {Sciences}} 119(48):e2216035119, \urlprefix\url{http://dx.doi.org/10.1073/pnas.2216035119}, publisher: Proceedings of the National Academy of Sciences.

\bibitem[{Bonacich(1987)}]{bonacich1987power}
Bonacich P (1987) {Power} {And} {Centrality}: A {Family} {Of} {Measures}. \emph{{American} {Journal} {Of} {Sociology}} 92(5):1170--1182.

\bibitem[{Brandes et~al.(2006)Brandes, Delling, Gaertler, G{\"o}rke, Hoefer, Nikoloski, \protect\BIBand{} Wagner}]{brandes2006maximizing}
Brandes U, Delling D, Gaertler M, G{\"o}rke R, Hoefer M, Nikoloski Z, Wagner D (2006) {Maximizing} {Modularity} {Is} {Hard}. \emph{{Arxiv} {Preprint} {Physics}/0608255} .

\bibitem[{Brown et~al.(2020)Brown, Mann, Ryder, Subbiah, Kaplan, Dhariwal, Neelakantan, Shyam, Sastry, Askell et~al.}]{brown2020language}
Brown T, Mann B, Ryder N, Subbiah M, Kaplan JD, Dhariwal P, Neelakantan A, Shyam P, Sastry G, Askell A, et~al. (2020) {Language} {Models} {Are} {Few}-{Shot} {Learners}. \emph{{Advances} {In} {Neural} {Information} {Processing} {Systems}} 33:1877--1901.

\bibitem[{Chen et~al.(2010)Chen, Wang, \protect\BIBand{} Wang}]{chen_scalable_2010}
Chen W, Wang C, Wang Y (2010) {Scalable} {Influence} {Maximization} {For} {Prevalent} {Viral} {Marketing} {In} {Large}-{Scale} {Social} {Networks}. \emph{Proceedings of the 16th {ACM} {SIGKDD} international conference on {Knowledge} discovery and data mining}, 1029--1038, {KDD} '10 (New York, NY, USA: Association for Computing Machinery), ISBN 978-1-4503-0055-1, \urlprefix\url{http://dx.doi.org/10.1145/1835804.1835934}.

\bibitem[{Chen et~al.(2009)Chen, Wang, \protect\BIBand{} Yang}]{chen2009efficient}
Chen W, Wang Y, Yang S (2009) {Efficient} {Influence} {Maximization} {In} {Social} {Networks}. \emph{{Proceedings} {Of} {The} 15Th {ACM} {SIGKDD} {International} {Conference} {On} {Knowledge} {Discovery} {And} {Data} {Mining}}, 199--208.

\bibitem[{Chen \protect\BIBand{} Zaman(2024)}]{chen2024shadowban}
Chen YS, Zaman T (2024) {Shaping} {Opinions} in {Social} {Networks} with {Shadow} {Banning}. \emph{{Plos} {One}} 19(3):e0299977.

\bibitem[{Chen \protect\BIBand{} Zaman(2025)}]{chen2025optimizing}
Chen YS, Zaman T (2025) {Optimizing} {Influence} {Campaigns}: {Nudging} under {Bounded} {Confidence}. \emph{{Arxiv} {Preprint} {Arxiv}:2503.18331} .

\bibitem[{Cinus et~al.(2025)Cinus, Minici, Luceri, \protect\BIBand{} Ferrara}]{cinus_exposing_2025}
Cinus F, Minici M, Luceri L, Ferrara E (2025) Exposing {Cross}-{Platform} {Coordinated} {Inauthentic} {Activity} in the {Run}-{Up} to the 2024 {U}.{S}. {Election}. \emph{Proceedings of the {ACM} on {Web} {Conference} 2025}, 541--559, {WWW} '25 (New York, NY, USA: Association for Computing Machinery), ISBN 9798400712746, \urlprefix\url{http://dx.doi.org/10.1145/3696410.3714698}.

\bibitem[{Compton et~al.(2014)Compton, Jurgens, \protect\BIBand{} Allen}]{geotagging}
Compton R, Jurgens D, Allen D (2014) {Geotagging} {One} {Hundred} {Million} {Twitter} {Accounts} {With} {Total} {Variation} {Minimization}. \emph{2014 {IEEE} {International} {Conference} {On} {Big} {Data} ({Big} {Data})}, 393--401 (IEEE).

\bibitem[{Costello et~al.(2024)Costello, Pennycook, \protect\BIBand{} Rand}]{costello2024durably}
Costello TH, Pennycook G, Rand DG (2024) {Durably} {Reducing} {Conspiracy} {Beliefs} {Through} {Dialogues} {With} {AI}. \emph{{Science}} 385(6714):eadq1814.

\bibitem[{Davis et~al.(2016)Davis, Varol, Ferrara, Flammini, \protect\BIBand{} Menczer}]{davis2016botornot}
Davis CA, Varol O, Ferrara E, Flammini A, Menczer F (2016) {Botornot}: A {System} {To} {Evaluate} {Social} {Bots}. \emph{{Proceedings} {Of} {The} 25Th {International} {Conference} {Companion} {On} {World} {Wide} {Web}}, 273--274.

\bibitem[{de~Vos et~al.(2024)de~Vos, Borm, \protect\BIBand{} Hamers}]{de_vos_influencing_2024}
de~Vos W, Borm P, Hamers H (2024) Influencing {Opinion} {Networks}: {Optimization} and {Games}. \emph{{Dynamic} {Games} {And} {Applications}} 14(4):959--980, ISSN 2153-0793, \urlprefix\url{http://dx.doi.org/10.1007/s13235-023-00543-6}.

\bibitem[{Deffuant et~al.(2001)Deffuant, Neau, Amblard, \protect\BIBand{} Weisbuch}]{deffuant_mixing_2001}
Deffuant G, Neau D, Amblard F, Weisbuch G (2001) {Mixing} {Beliefs} {Among} {Interacting} {Agents}. \emph{{Advances} {In} {Complex} {Systems}} 3:87--98.

\bibitem[{DeGroot(1974)}]{degroot1974reaching}
DeGroot MH (1974) {Reaching} a {Consensus}. \emph{{Journal} {Of} {The} {American} {Statistical} {Association}} 69(345):118--121.

\bibitem[{des Mesnards et~al.(2022)des Mesnards, Hunter, el~Hjouji, \protect\BIBand{} Zaman}]{des2022detecting}
des Mesnards NG, Hunter DS, el~Hjouji Z, Zaman T (2022) {Detecting} {Bots} {And} {Assessing} {Their} {Impact} {In} {Social} {Networks}. \emph{{Operations} {Research}} 70(1):1--22.

\bibitem[{Devlin et~al.(2018)Devlin, Chang, Lee, \protect\BIBand{} Toutanova}]{devlin2018bert}
Devlin J, Chang MW, Lee K, Toutanova K (2018) {BERT}: {Pre}-{Training} {Of} {Deep} {Bidirectional} {Transformers} {For} {Language} {Understanding}. \emph{{Arxiv} {Preprint} {Arxiv}:1810.04805} .

\bibitem[{Drenten \protect\BIBand{} Brooks(2020)}]{drenten2020celebrity}
Drenten J, Brooks G (2020) {Celebrity} 2.0: {Lil} {Miquela} {And} {The} {Rise} {Of} a {Virtual} {Star} {System}. \emph{{Feminist} {Media} {Studies}} 20(8):1319--1323.

\bibitem[{Du et~al.(2013)Du, Song, Gomez-Rodriguez, \protect\BIBand{} Zha}]{du_scalable_2013}
Du N, Song L, Gomez-Rodriguez M, Zha H (2013) {Scalable} {Influence} {Estimation} {In} {Continuous}-{Time} {Diffusion} {Networks}. \emph{Proceedings of the 27th {International} {Conference} on {Neural} {Information} {Processing} {Systems} - {Volume} 2}, volume~2 of \emph{{NIPS}'13}, 3147--3155 (Red Hook, NY, USA: Curran Associates Inc.).

\bibitem[{Elyassami et~al.(2021)Elyassami, Alseiari, ALZaabi, Hashem, \protect\BIBand{} Aljahoori}]{elyassami2021fake}
Elyassami S, Alseiari S, ALZaabi M, Hashem A, Aljahoori N (2021) {Fake} {News} {Detection} {Using} {Ensemble} {Learning} {And} {Machine} {Learning} {Algorithms}. \emph{{Combating} {Fake} {News} {With} {Computational} {Intelligence} {Techniques}}, 149--162 (Springer).

\bibitem[{Farajtabar et~al.(2014)Farajtabar, Du, Gomez-Rodriguez, Valera, Zha, \protect\BIBand{} Song}]{farajtabar_shaping_2014}
Farajtabar M, Du N, Gomez-Rodriguez M, Valera I, Zha H, Song L (2014) Shaping {Social} {Activity} by {Incentivizing} {Users}. \emph{Advances in {Neural} {Information} {Processing} {Systems}}, volume~27 (Curran Associates, Inc.), \urlprefix\url{https://papers.nips.cc/paper_files/paper/2014/hash/5bce7b2ded63483e6f2828e6a9427498-Abstract.html}.

\bibitem[{Farajtabar et~al.(2017)Farajtabar, Yang, Ye, Xu, Trivedi, Khalil, Li, Song, \protect\BIBand{} Zha}]{farajtabar_fake_2017}
Farajtabar M, Yang J, Ye X, Xu H, Trivedi R, Khalil E, Li S, Song L, Zha H (2017) {Fake} {News} {Mitigation} {Via} {Point} {Process} {Based} {Intervention}. \emph{Proceedings of the 34th {International} {Conference} on {Machine} {Learning} - {Volume} 70}, 1097--1106, {ICML}'17 (Sydney, NSW, Australia: JMLR.org).

\bibitem[{Ferrara et~al.(2016)Ferrara, Varol, Davis, Menczer, \protect\BIBand{} Flammini}]{ferrara2016rise}
Ferrara E, Varol O, Davis C, Menczer F, Flammini A (2016) {The} {Rise} {Of} {Social} {Bots}. \emph{{Communications} {Of} {The} {ACM}} 59(7):96--104.

\bibitem[{Festinger(1954)}]{festinger1954theory}
Festinger L (1954) A {Theory} {Of} {Social} {Comparison} {Processes}. \emph{{Human} {Relations}} 7(2):117--140.

\bibitem[{Fortunato(2010)}]{fortunato2010community}
Fortunato S (2010) {Community} {Detection} {In} {Graphs}. \emph{{Physics} {Reports}} 486(3--5):75--174.

\bibitem[{François(2020)}]{francois_actors_2020}
François C (2020) Actors, {Behaviors}, {Content}: {A} {Disinformation} {ABC}. \emph{{Algorithms}} \urlprefix\url{https://www.annenbergpublicpolicycenter.org/wp-content/uploads/ABC_Framework_TWG_Francois_Sept_2019.pdf}.

\bibitem[{Freeman(1977)}]{freeman1977set}
Freeman LC (1977) A {Set} {Of} {Measures} {Of} {Centrality} {Based} {On} {Betweenness}. \emph{{Sociometry}} 35--41.

\bibitem[{Galasso et~al.(2023)Galasso, Nannicini, \protect\BIBand{} Nunnari}]{galasso_positive_2023}
Galasso V, Nannicini T, Nunnari S (2023) Positive {Spillovers} from {Negative} {Campaigning}. \emph{{American} {Journal} {Of} {Political} {Science}} 67(1):5--21, ISSN 1540-5907, \urlprefix\url{http://dx.doi.org/10.1111/ajps.12610}, \_eprint: https://onlinelibrary.wiley.com/doi/pdf/10.1111/ajps.12610.

\bibitem[{Garimella et~al.(2018)Garimella, De~Francisci~Morales, Gionis, \protect\BIBand{} Mathioudakis}]{garimella2018political}
Garimella K, De~Francisci~Morales G, Gionis A, Mathioudakis M (2018) {Political} {Discourse} {On} {Social} {Media}: {Echo} {Chambers}, {Gatekeepers}, {And} {The} {Price} {Of} {Bipartisanship}. \emph{{Proceedings} {Of} {The} 2018 {World} {Wide} {Web} {Conference}}, 913--922.

\bibitem[{Ghaderi \protect\BIBand{} Srikant(2013)}]{ghaderi2013opinion}
Ghaderi J, Srikant R (2013) {Opinion} {Dynamics} {In} {Social} {Networks}: A {Local} {Interaction} {Game} {With} {Stubborn} {Agents}. \emph{{American} {Control} {Conference} ({Acc}), 2013}, 1982--1987 (IEEE).

\bibitem[{Goldstein et~al.(2023)Goldstein, Sastry, Musser, DiResta, Gentzel, \protect\BIBand{} Sedova}]{goldstein_generative_2023}
Goldstein JA, Sastry G, Musser M, DiResta R, Gentzel M, Sedova K (2023) Generative {Language} {Models} and {Automated} {Influence} {Operations}: {Emerging} {Threats} and {Potential} {Mitigations}. \urlprefix\url{http://dx.doi.org/10.48550/arXiv.2301.04246}, arXiv:2301.04246 [cs].

\bibitem[{Guess et~al.(2023)Guess, Malhotra, Pan, Barber{\'a}, Allcott, Brown, Crespo-Tenorio, Dimmery, Freelon, Gentzkow et~al.}]{guess2023social}
Guess AM, Malhotra N, Pan J, Barber{\'a} P, Allcott H, Brown T, Crespo-Tenorio A, Dimmery D, Freelon D, Gentzkow M, et~al. (2023) {How} {Do} {Social} {Media} {Feed} {Algorithms} {Affect} {Attitudes} {And} {Behavior} {In} {An} {Election} {Campaign}? \emph{{Science}} 381(6656):398--404.

\bibitem[{Hagen \protect\BIBand{} Kahng(1992)}]{hagen1992new}
Hagen L, Kahng AB (1992) {New} {Spectral} {Methods} {For} {Ratio} {Cut} {Partitioning} {And} {Clustering}. \emph{{IEEE} {Transactions} {On} {Computer}-{Aided} {Design} {Of} {Integrated} {Circuits} {And} {Systems}} 11(9):1074--1085.

\bibitem[{Hegselmann et~al.(2015)Hegselmann, K{\"o}nig, Kurz, Niemann, \protect\BIBand{} Rambau}]{hegselmann2015optimal}
Hegselmann R, K{\"o}nig S, Kurz S, Niemann C, Rambau J (2015) {Optimal} {Opinion} {Control}: {The} {Campaign} {Problem}. \emph{{Journal} {Of} {Artificial} {Societies} {And} {Social} {Simulation}} 18(3):1--18.

\bibitem[{Hegselmann \protect\BIBand{} Krause(2002)}]{hegselmann2002opinion}
Hegselmann R, Krause U (2002) {Opinion} {Dynamics} {And} {Bounded} {Confidence}: {Models}, {Analysis} {And} {Simulation}. \emph{{Journal} {Of} {Artificial} {Societies} {And} {Social} {Simulation}} 5(3):1--33.

\bibitem[{Hirsch(2005)}]{hirsch2005index}
Hirsch JE (2005) {An} {Index} {To} {Quantify} {An} {Individual}'s {Scientific} {Research} {Output}. \emph{{Proceedings} {Of} {The} {National} {Academy} {Of} {Sciences}} 102(46):16569--16572.

\bibitem[{Hunter \protect\BIBand{} Zaman(2022)}]{hunter2022optimizing}
Hunter DS, Zaman T (2022) {Optimizing} {Opinions} {With} {Stubborn} {Agents}. \emph{{Operations} {Research}} 70(4):2119--2137.

\bibitem[{Hutto \protect\BIBand{} Gilbert(2014)}]{hutto2014vader}
Hutto C, Gilbert E (2014) {Vader}: A {Parsimonious} {Rule}-{Based} {Model} {For} {Sentiment} {Analysis} {Of} {Social} {Media} {Text}. \emph{{Eighth} {International} {Aaai} {Conference} {On} {Weblogs} {And} {Social} {Media}}.

\bibitem[{Iyengar \protect\BIBand{} Hahn(2009)}]{iyengar2009red}
Iyengar S, Hahn KS (2009) {Red} {Media}, {Blue} {Media}: {Evidence} {Of} {Ideological} {Selectivity} {In} {Media} {Use}. \emph{{Journal} {Of} {Communication}} 59(1):19--39.

\bibitem[{Jurafsky et~al.(2014)Jurafsky, Chahuneau, Routledge, \protect\BIBand{} Smith}]{jurafsky_narrative_2014}
Jurafsky D, Chahuneau V, Routledge BR, Smith NA (2014) {Narrative} {Framing} {Of} {Consumer} {Sentiment} {In} {Online} {Restaurant} {Reviews}. \emph{{First} {Monday}} ISSN 1396-0466, \urlprefix\url{http://dx.doi.org/10.5210/fm.v19i4.4944}.

\bibitem[{Juul \protect\BIBand{} Ugander(2021)}]{juul2021comparing}
Juul J, Ugander J (2021) {Comparing} {Information} {Diffusion} {Mechanisms} {By} {Matching} {On} {Cascade} {Size}. \emph{{Proceedings} {Of} {The} {National} {Academy} {Of} {Sciences}} 118(46):e2100786118.

\bibitem[{Kempe et~al.(2003)Kempe, Kleinberg, \protect\BIBand{} Tardos}]{kempe2003maximizing}
Kempe D, Kleinberg J, Tardos {\'E} (2003) {Maximizing} {The} {Spread} {Of} {Influence} {Through} a {Social} {Network}. \emph{{Proceedings} {Of} {The} {Ninth} {ACM} {SIGKDD} {International} {Conference} {On} {Knowledge} {Discovery} {And} {Data} {Mining}}, 137--146 (ACM).

\bibitem[{Klausen et~al.(2018)Klausen, Marks, \protect\BIBand{} Zaman}]{klausen2018finding}
Klausen J, Marks CE, Zaman T (2018) {Finding} {Extremists} {In} {Online} {Social} {Networks}. \emph{{Operations} {Research}} 66(4):957--976.

\bibitem[{Kong et~al.(2023)Kong, Calderon, Ram, Boichak, \protect\BIBand{} Rizoiu}]{kong_interval-censored_2023}
Kong Q, Calderon P, Ram R, Boichak O, Rizoiu MA (2023) Interval-censored {Transformer} {Hawkes}: {Detecting} {Information} {Operations} using the {Reaction} of {Social} {Systems}. \emph{Proceedings of the {ACM} {Web} {Conference} 2023}, 1813--1821, {WWW} '23 (New York, NY, USA: Association for Computing Machinery), ISBN 978-1-4503-9416-1, \urlprefix\url{http://dx.doi.org/10.1145/3543507.3583481}.

\bibitem[{Kozitsin(2022)}]{kozitsin_formal_2022}
Kozitsin IV (2022) {Formal} {Models} {Of} {Opinion} {Formation} {And} {Their} {Application} {To} {Real} {Data}: {Evidence} {From} {Online} {Social} {Networks}. \emph{{The} {Journal} {Of} {Mathematical} {Sociology}} 46(2):120--147, ISSN 0022-250X, \urlprefix\url{http://dx.doi.org/10.1080/0022250X.2020.1835894}, place: PHILADELPHIA Publisher: Taylor \& Francis.

\bibitem[{Kozyreva et~al.(2024)Kozyreva, Lorenz-Spreen, Herzog, Ecker, Lewandowsky, Hertwig, Ali, Bak-Coleman, Barzilai, Basol et~al.}]{kozyreva2024toolbox}
Kozyreva A, Lorenz-Spreen P, Herzog SM, Ecker UK, Lewandowsky S, Hertwig R, Ali A, Bak-Coleman J, Barzilai S, Basol M, et~al. (2024) {Toolbox} {Of} {Individual}-{Level} {Interventions} {Against} {Online} {Misinformation}. \emph{{Nature} {Human} {Behaviour}} 8(6):1044--1052.

\bibitem[{Kunegis(2013)}]{konect}
Kunegis J (2013) {Konect}: {The} {Koblenz} {Network} {Collection}. \emph{{Proceedings} {Of} {The} 22Nd {International} {Conference} {On} {World} {Wide} {Web}}, 1343--1350 (ACM).

\bibitem[{Laszkiewicz \protect\BIBand{} Kalinska-Kula(2023)}]{laszkiewicz2023virtual}
Laszkiewicz A, Kalinska-Kula M (2023) {Virtual} {Influencers} {As} {An} {Emerging} {Marketing} {Theory}: A {Systematic} {Literature} {Review}. \emph{{International} {Journal} {Of} {Consumer} {Studies}} 47(6):2479--2494.

\bibitem[{Lei et~al.(2015)Lei, Maniu, Mo, Cheng, \protect\BIBand{} Senellart}]{lei_online_2015}
Lei S, Maniu S, Mo L, Cheng R, Senellart P (2015) Online {Influence} {Maximization}. \emph{Proceedings of the 21th {ACM} {SIGKDD} {International} {Conference} on {Knowledge} {Discovery} and {Data} {Mining}}, 645--654, {KDD} '15 (New York, NY, USA: Association for Computing Machinery), ISBN 978-1-4503-3664-2, \urlprefix\url{http://dx.doi.org/10.1145/2783258.2783271}.

\bibitem[{Leskovec et~al.(2007)Leskovec, Adamic, \protect\BIBand{} Huberman}]{leskovec_dynamics_2007}
Leskovec J, Adamic LA, Huberman BA (2007) {The} {Dynamics} {Of} {Viral} {Marketing}. \emph{{Acm} {Trans}. {Web}} 1(1):5--es, ISSN 1559-1131, \urlprefix\url{http://dx.doi.org/10.1145/1232722.1232727}.

\bibitem[{Leskovec \protect\BIBand{} Krevl(2014)}]{snap_datasets}
Leskovec J, Krevl A (2014) {SNAP} {Datasets}: {Stanford} {Large} {Network} {Dataset} {Collection}. \url{http://snap.stanford.edu/data}.

\bibitem[{Lindelauf et~al.(2011)Lindelauf, Borm, \protect\BIBand{} Hamers}]{lindelauf_understanding_2011}
Lindelauf R, Borm P, Hamers H (2011) Understanding {Terrorist} {Network} {Topologies} and {Their} {Resilience} {Against} {Disruption}. Wiil UK, ed., \emph{Counterterrorism and {Open} {Source} {Intelligence}}, 61--72 (Vienna: Springer), ISBN 978-3-7091-0388-3, \urlprefix\url{http://dx.doi.org/10.1007/978-3-7091-0388-3_5}.

\bibitem[{Lindelauf et~al.(2013)Lindelauf, Hamers, \protect\BIBand{} Husslage}]{lindelauf_cooperative_2013}
Lindelauf R, Hamers H, Husslage B (2013) Cooperative game theoretic centrality analysis of terrorist networks: {The} cases of {Jemaah} {Islamiyah} and {Al} {Qaeda}. \emph{{European} {Journal} {Of} {Operational} {Research}} 229(1):230--238, ISSN 03772217, \urlprefix\url{http://dx.doi.org/10.1016/j.ejor.2013.02.032}.

\bibitem[{Lorenz(2006)}]{lorenz2006consensus}
Lorenz J (2006) {Consensus} {Strikes} {Back} {In} {The} {Hegselmann}–{Krause} {Model} {Of} {Continuous} {Opinion} {Dynamics} {Under} {Bounded} {Confidence}. \emph{{Journal} {Of} {Artificial} {Societies} {And} {Social} {Simulation}} 9(1).

\bibitem[{Luceri et~al.(2024{\natexlab{a}})Luceri, Boniardi, \protect\BIBand{} Ferrara}]{luceri_leveraging_2024}
Luceri L, Boniardi E, Ferrara E (2024{\natexlab{a}}) Leveraging {Large} {Language} {Models} to {Detect} {Influence} {Campaigns} on {Social} {Media}. \emph{Companion {Proceedings} of the {ACM} {Web} {Conference} 2024}, 1459--1467, {WWW} '24 (New York, NY, USA: Association for Computing Machinery), ISBN 9798400701726, \urlprefix\url{http://dx.doi.org/10.1145/3589335.3651912}.

\bibitem[{Luceri et~al.(2024{\natexlab{b}})Luceri, Pantè, Burghardt, \protect\BIBand{} Ferrara}]{luceri_unmasking_2024}
Luceri L, Pantè V, Burghardt K, Ferrara E (2024{\natexlab{b}}) Unmasking the {Web} of {Deceit}: {Uncovering} {Coordinated} {Activity} to {Expose} {Information} {Operations} on {Twitter}. \emph{Proceedings of the {ACM} {Web} {Conference} 2024}, 2530--2541, {WWW} '24 (New York, NY, USA: Association for Computing Machinery), ISBN 9798400701719, \urlprefix\url{http://dx.doi.org/10.1145/3589334.3645529}.

\bibitem[{Lyu \protect\BIBand{} Lo(2020)}]{lyu2020fake}
Lyu S, Lo DCT (2020) {Fake} {News} {Detection} {By} {Decision} {Tree}. \emph{2020 {Southeastcon}}, 1--2 (IEEE).

\bibitem[{Maier et~al.(2018)Maier, ~, ~, ~, ~, ~, ~, ~, ~, ~, ~, , \protect\BIBand{} Adam}]{maier_applying_2018}
Maier D, ~ W A, ~ M P, ~ W G, ~ N A, ~ K A, ~ P B, ~ H G, ~ R U, ~ H T, ~ SP H, , Adam S (2018) Applying {LDA} {Topic} {Modeling} in {Communication} {Research}: {Toward} a {Valid} and {Reliable} {Methodology}. \emph{{Communication} {Methods} {And} {Measures}} 12(2-3):93--118, ISSN 1931-2458, \urlprefix\url{http://dx.doi.org/10.1080/19312458.2018.1430754}, publisher: Routledge \_eprint: https://doi.org/10.1080/19312458.2018.1430754.

\bibitem[{Marks \protect\BIBand{} Zaman(2022)}]{marks2022building}
Marks CE, Zaman T (2022) {Building} a {Location}-{Based} {Set} {Of} {Social} {Media} {Users}. \emph{{Operations} {Research}} 70(6):3090--3107.

\bibitem[{Marks \protect\BIBand{} Tegmark(2023)}]{marks2023geometry}
Marks S, Tegmark M (2023) {The} {Geometry} {Of} {Truth}: {Emergent} {Linear} {Structure} {In} {Large} {Language} {Model} {Representations} {Of} {True}/{False} {Datasets}. \emph{{Arxiv} {Preprint} {Arxiv}:2310.06824} .

\bibitem[{Masuda(2015)}]{masuda2015opinion}
Masuda N (2015) {Opinion} {Control} {In} {Complex} {Networks}. \emph{{New} {Journal} {Of} {Physics}} 17(3):033031.

\bibitem[{McPherson et~al.(2001)McPherson, Smith-Lovin, \protect\BIBand{} Cook}]{mcpherson2001birds}
McPherson M, Smith-Lovin L, Cook JM (2001) {Birds} {Of} a {Feather}: {Homophily} {In} {Social} {Networks}. \emph{{Annual} {Review} {Of} {Sociology}} 27:415--444.

\bibitem[{Mesnards et~al.(2018)Mesnards, Hunter, Hjouji, \protect\BIBand{} Zaman}]{mesnards2018detecting}
Mesnards NGd, Hunter DS, Hjouji Ze, Zaman T (2018) {Detecting} {Bots} {And} {Assessing} {Their} {Impact} {In} {Social} {Networks}. \emph{{Arxiv} {Preprint} {Arxiv}:1810.12398} .

\bibitem[{Minici et~al.(2025)Minici, Luceri, Fabbri, \protect\BIBand{} Ferrara}]{minici_iohunter_2025}
Minici M, Luceri L, Fabbri F, Ferrara E (2025) {IOHunter}: {Graph} {Foundation} {Model} to {Uncover} {Online} {Information} {Operations}. \emph{{Proceedings} {Of} {The} {Aaai} {Conference} {On} {Artificial} {Intelligence}} 39(27):28258--28266, ISSN 2374-3468, \urlprefix\url{http://dx.doi.org/10.1609/aaai.v39i27.35046}, number: 27.

\bibitem[{Mobilia et~al.(2007)Mobilia, Petersen, \protect\BIBand{} Redner}]{mobilia2007role}
Mobilia M, Petersen AE, Redner S (2007) {Role} {Of} {Zealotry} {In} {The} {Voter} {Model}. \emph{{Journal} {Of} {Statistical} {Mechanics}: {Theory} {And} {Experiment}} 2007(08):P08029.

\bibitem[{Monti et~al.(2019)Monti, Frasca, Eynard, Mannion, \protect\BIBand{} Bronstein}]{monti2019fake}
Monti F, Frasca F, Eynard D, Mannion D, Bronstein MM (2019) {Fake} {News} {Detection} {On} {Social} {Media} {Using} {Geometric} {Deep} {Learning}. \emph{{Arxiv} {Preprint} {Arxiv}:1902.06673} .

\bibitem[{Mueller et~al.(2019)}]{mueller2019mueller}
Mueller RS, et~al. (2019) \emph{{The} {Mueller} {Report}: {Complete} {Report} {On} {The} {Investigation} {Into} {Russian} {Interference} {In} {The} 2016 {Presidential} {Election}} (e-artnow).

\bibitem[{Munger(2017)}]{munger2017tweetment}
Munger K (2017) {Tweetment} {Effects} {On} {The} {Tweeted}: {Experimentally} {Reducing} {Racist} {Harassment}. \emph{{Political} {Behavior}} 39:629--649.

\bibitem[{Newman(2006)}]{newman2006modularity}
Newman ME (2006) {Modularity} {And} {Community} {Structure} {In} {Networks}. \emph{{Proceedings} {Of} {The} {National} {Academy} {Of} {Sciences}} 103(23):8577--8582.

\bibitem[{Ng et~al.(2001)Ng, Jordan, \protect\BIBand{} Weiss}]{ng2001spectral}
Ng A, Jordan M, Weiss Y (2001) {On} {Spectral} {Clustering}: {Analysis} {And} {An} {Algorithm}. \emph{{Advances} {In} {Neural} {Information} {Processing} {Systems}} 14.

\bibitem[{Nickerson(1998)}]{nickerson1998confirmation}
Nickerson RS (1998) {Confirmation} {Bias}: A {Ubiquitous} {Phenomenon} {In} {Many} {Guises}. \emph{{Review} {Of} {General} {Psychology}} 2(2):175--220.

\bibitem[{Nyhan \protect\BIBand{} Reifler(2010)}]{nyhan_when_2010}
Nyhan B, Reifler J (2010) When {Corrections} {Fail}: {The} {Persistence} of {Political} {Misperceptions}. \emph{{Political} {Behavior}} 32(2):303--330, ISSN 1573-6687, \urlprefix\url{http://dx.doi.org/10.1007/s11109-010-9112-2}.

\bibitem[{Nyhan et~al.(2023)Nyhan, Settle, Thorson, Wojcieszak, Barber{\'a}, Chen, Allcott, Brown, Crespo-Tenorio, Dimmery et~al.}]{nyhan2023like}
Nyhan B, Settle J, Thorson E, Wojcieszak M, Barber{\'a} P, Chen AY, Allcott H, Brown T, Crespo-Tenorio A, Dimmery D, et~al. (2023) {Like}-{Minded} {Sources} {On} {Facebook} {Are} {Prevalent} {But} {Not} {Polarizing}. \emph{{Nature}} 620(7972):137--144.

\bibitem[{{OpenAI}(2022)}]{openai_chatgpt}
{OpenAI} (2022) {Introducing} {Chatgpt}. \url{https://openai.com/index/chatgpt/}, \urlprefix\url{https://openai.com/index/chatgpt/}, accessed: 2025-04-13.

\bibitem[{Padgett \protect\BIBand{} Ansell(1993)}]{padgett1993robust}
Padgett JF, Ansell CK (1993) {Robust} {Action} {And} {The} {Rise} {Of} {The} {Medici}, 1400–1434. \emph{{American} {Journal} {Of} {Sociology}} 98(6):1259--1319.

\bibitem[{Page et~al.(1999)Page, Brin, Motwani, \protect\BIBand{} Winograd}]{page1999pagerank}
Page L, Brin S, Motwani R, Winograd T (1999) {The} {Pagerank} {Citation} {Ranking}: {Bringing} {Order} {To} {The} {Web}. \emph{{Stanford} {Infolab}} .

\bibitem[{Pamment(2020)}]{pamment_abcde_2020}
Pamment J (2020) The {ABCDE} {Framework}. Technical report, Carnegie Endowment for International Peace, \urlprefix\url{https://www.jstor.org/stable/resrep26180.6}.

\bibitem[{Park et~al.(2023)Park, O'Brien, Cai, Morris, Liang, \protect\BIBand{} Bernstein}]{park2023generative}
Park JS, O'Brien J, Cai CJ, Morris MR, Liang P, Bernstein MS (2023) {Generative} {Agents}: {Interactive} {Simulacra} {Of} {Human} {Behavior}. \emph{{Proceedings} {Of} {The} 36Th {Annual} {Acm} {Symposium} {On} {User} {Interface} {Software} {And} {Technology}}, 1--22.

\bibitem[{Park et~al.(2024)Park, Zou, Shaw, Hill, Cai, Morris, Willer, Liang, \protect\BIBand{} Bernstein}]{llm_persona_simulation}
Park JS, Zou CQ, Shaw A, Hill BM, Cai C, Morris MR, Willer R, Liang P, Bernstein MS (2024) {Generative} {Agent} {Simulations} {Of} 1,000 {People}. \emph{{Arxiv} {Preprint} {Arxiv}:2411.10109} .

\bibitem[{Pennycook et~al.(2021)Pennycook, Epstein, Mosleh, Arechar, Eckles, \protect\BIBand{} Rand}]{pennycook2021shifting}
Pennycook G, Epstein Z, Mosleh M, Arechar AA, Eckles D, Rand DG (2021) {Shifting} {Attention} {To} {Accuracy} {Can} {Reduce} {Misinformation} {Online}. \emph{{Nature}} 592(7855):590--595.

\bibitem[{Pennycook \protect\BIBand{} Rand(2019)}]{pennycook2019fighting}
Pennycook G, Rand DG (2019) {Fighting} {Misinformation} {On} {Social} {Media} {Using} {Crowdsourced} {Judgments} {Of} {News} {Source} {Quality}. \emph{{Proceedings} {Of} {The} {National} {Academy} {Of} {Sciences}} 116(7):2521--2526.

\bibitem[{Pr{\"o}llochs(2022)}]{prollochs2022community}
Pr{\"o}llochs N (2022) {Community}-{Based} {Fact}-{Checking} {On} {Twitter}’s {Birdwatch} {Platform}. \emph{{Proceedings} {Of} {The} {International} {AAAI} {Conference} {On} {Web} {And} {Social} {Media}}, volume~16, 794--805.

\bibitem[{Radford et~al.(2021)Radford, Kim, Hallacy, Ramesh, Goh, Agarwal, Sastry, Askell, Mishkin, Clark et~al.}]{radford2021learning}
Radford A, Kim JW, Hallacy C, Ramesh A, Goh G, Agarwal S, Sastry G, Askell A, Mishkin P, Clark J, et~al. (2021) {Learning} {Transferable} {Visual} {Models} {From} {Natural} {Language} {Supervision}. \emph{{International} {Conference} {On} {Machine} {Learning}}, 8748--8763 (PmLR).

\bibitem[{Radford et~al.(2018)Radford, Narasimhan, Salimans, Sutskever et~al.}]{radford2018improving}
Radford A, Narasimhan K, Salimans T, Sutskever I, et~al. (2018) {Improving} {Language} {Understanding} {By} {Generative} {Pre}-{Training}. \url{https://cdn.openai.com/research-covers/language-unsupervised/language_understanding_paper.pdf}.

\bibitem[{Ramesh et~al.(2022)Ramesh, Dhariwal, Nichol, Chu, \protect\BIBand{} Chen}]{ramesh2022hierarchical}
Ramesh A, Dhariwal P, Nichol A, Chu C, Chen M (2022) {Hierarchical} {Text}-{Conditional} {Image} {Generation} {With} {Clip} {Latents}. \emph{{Arxiv} {Preprint} {Arxiv}:2204.06125} 1(2):3.

\bibitem[{Ramesh et~al.(2021)Ramesh, Pavlov, Goh, Gray, Voss, Radford, Chen, \protect\BIBand{} Sutskever}]{ramesh2021zero}
Ramesh A, Pavlov M, Goh G, Gray S, Voss C, Radford A, Chen M, Sutskever I (2021) {Zero}-{Shot} {Text}-{To}-{Image} {Generation}. \emph{{International} {Conference} {On} {Machine} {Learning}}, 8821--8831 (Pmlr).

\bibitem[{Ribeiro et~al.(2020)Ribeiro, Ottoni, West, Almeida, \protect\BIBand{} Meira}]{ribeiro_auditing_2020}
Ribeiro MH, Ottoni R, West R, Almeida VAF, Meira W (2020) Auditing radicalization pathways on {YouTube}. \emph{Proceedings of the 2020 {Conference} on {Fairness}, {Accountability}, and {Transparency}}, 131--141, {FAT}* '20 (New York, NY, USA: Association for Computing Machinery), ISBN 978-1-4503-6936-7, \urlprefix\url{http://dx.doi.org/10.1145/3351095.3372879}.

\bibitem[{Rizoiu et~al.(2017)Rizoiu, Lee, Mishra, Xie, Caetano, \protect\BIBand{} Lepri}]{rizoiu2017hawkes}
Rizoiu MA, Lee Y, Mishra S, Xie L, Caetano TS, Lepri B (2017) A {Tutorial} {On} {Hawkes} {Processes} {For} {Events} {In} {Social} {Media}. \emph{{Arxiv} {Preprint} {Arxiv}:1708.06401} .

\bibitem[{Rocklage et~al.(2023)Rocklage, He, Rucker, \protect\BIBand{} Nordgren}]{rocklage_beyond_2023}
Rocklage MD, He S, Rucker DD, Nordgren LF (2023) Beyond {Sentiment}: {The} {Value} and {Measurement} of {Consumer} {Certainty} in {Language}. \emph{{Journal} {Of} {Marketing} {Research}} 60(5):870--888, ISSN 0022-2437, \urlprefix\url{http://dx.doi.org/10.1177/00222437221134802}, publisher: SAGE Publications Inc.

\bibitem[{Rossetti \protect\BIBand{} Zaman(2023)}]{rossetti2023bots}
Rossetti M, Zaman T (2023) {Bots}, {Disinformation}, {And} {The} {First} {Impeachment} {Of} {US} {President} {Donald} {Trump}. \emph{{Plos} {One}} 18(5):e0283971.

\bibitem[{Rossi \protect\BIBand{} Ahmed(2015)}]{network_repository}
Rossi R, Ahmed N (2015) The network data repository with interactive graph analytics and visualization. \emph{Proceedings of the AAAI conference on artificial intelligence}, volume~29.

\bibitem[{Sabidussi(1966)}]{sabidussi1966centrality}
Sabidussi G (1966) {The} {Centrality} {Index} {Of} a {Graph}. \emph{{Psychometrika}} 31(4):581--603.

\bibitem[{Schneider \protect\BIBand{} Rizoiu(2023)}]{schneider_effectiveness_2023}
Schneider PJ, Rizoiu MA (2023) {The} {Effectiveness} {Of} {Moderating} {Harmful} {Online} {Content}. \emph{{Proceedings} {Of} {The} {National} {Academy} {Of} {Sciences}} 120(34):e2307360120, \urlprefix\url{http://dx.doi.org/10.1073/pnas.2307360120}, publisher: Proceedings of the National Academy of Sciences.

\bibitem[{Seckin et~al.(2024)Seckin, Pote, Nwala, Yin, Luceri, Flammini, \protect\BIBand{} Menczer}]{seckin_labeled_2024}
Seckin OC, Pote M, Nwala A, Yin L, Luceri L, Flammini A, Menczer F (2024) Labeled {Datasets} for {Research} on {Information} {Operations}. \urlprefix\url{http://dx.doi.org/10.48550/arXiv.2411.10609}, arXiv:2411.10609 [cs].

\bibitem[{Shah \protect\BIBand{} Zaman(2011)}]{shah2011rumors}
Shah D, Zaman T (2011) {Rumors} {In} a {Network}: {Who}’s {The} {Culprit}? \emph{{IEEE} {Transactions} {On} {Information} {Theory}} 57(8):5163--5181.

\bibitem[{Shah \protect\BIBand{} Zaman(2016)}]{shah2016finding}
Shah D, Zaman T (2016) {Finding} {Rumor} {Sources} {On} {Random} {Trees}. \emph{{Operations} {Research}} 64(3):736--755.

\bibitem[{Shi \protect\BIBand{} Malik(2000)}]{shi2000normalized}
Shi J, Malik J (2000) {Normalized} {Cuts} {And} {Image} {Segmentation}. \emph{{IEEE} {Transactions} {On} {Pattern} {Analysis} {And} {Machine} {Intelligence}} 22(8):888--905.

\bibitem[{Shu et~al.(2017)Shu, Sliva, Wang, Tang, \protect\BIBand{} Liu}]{shu2017fake}
Shu K, Sliva A, Wang S, Tang J, Liu H (2017) {Fake} {News} {Detection} {On} {Social} {Media}: A {Data} {Mining} {Perspective}. \emph{{ACM} {SIGKDD} {Explorations} {Newsletter}} 19(1):22--36.

\bibitem[{Tessler et~al.(2024)Tessler, Bakker, Jarrett, Sheahan, Chadwick, Koster, Evans, Campbell-Gillingham, Collins, Parkes et~al.}]{tessler2024ai}
Tessler MH, Bakker MA, Jarrett D, Sheahan H, Chadwick MJ, Koster R, Evans G, Campbell-Gillingham L, Collins T, Parkes DC, et~al. (2024) {AI} {Can} {Help} {Humans} {Find} {Common} {Ground} {In} {Democratic} {Deliberation}. \emph{{Science}} 386(6719):eadq2852.

\bibitem[{Thaler \protect\BIBand{} Sunstein(2009)}]{thaler2009nudge}
Thaler RH, Sunstein CR (2009) \emph{{Nudge}: {Improving} {Decisions} about {Health}, {Wealth}, and {Happiness}} (Penguin).

\bibitem[{Trujillo et~al.(2025)Trujillo, Fagni, \protect\BIBand{} Cresci}]{trujillo_dsa_2025}
Trujillo A, Fagni T, Cresci S (2025) The {DSA} {Transparency} {Database}: {Auditing} {Self}-reported {Moderation} {Actions} by {Social} {Media}. \emph{{Proc}. {Acm} {Hum}.-{Comput}. {Interact}.} 9(2):CSCW187:1--CSCW187:28, \urlprefix\url{http://dx.doi.org/10.1145/3711085}.

\bibitem[{Vassio et~al.(2014)Vassio, Fagnani, Frasca, \protect\BIBand{} Ozdaglar}]{vassio2014message}
Vassio L, Fagnani F, Frasca P, Ozdaglar A (2014) {Message} {Passing} {Optimization} {Of} {Harmonic} {Influence} {Centrality}. \emph{{IEEE} {Transactions} {On} {Control} {Of} {Network} {Systems}} 1(1):109--120.

\bibitem[{Vaswani et~al.(2017)Vaswani, Shazeer, Parmar, Uszkoreit, Jones, Gomez, Kaiser, \protect\BIBand{} Polosukhin}]{vaswani2017attention}
Vaswani A, Shazeer N, Parmar N, Uszkoreit J, Jones L, Gomez AN, Kaiser L, Polosukhin I (2017) {Attention} {Is} {All} {You} {Need}. \emph{{Advances} {In} {Neural} {Information} {Processing} {Systems}}, volume~30.

\bibitem[{Von~Luxburg(2007)}]{von2007tutorial}
Von~Luxburg U (2007) A {Tutorial} {On} {Spectral} {Clustering}. \emph{{Statistics} {And} {Computing}} 17(4):395--416.

\bibitem[{Vosoughi et~al.(2018)Vosoughi, Roy, \protect\BIBand{} Aral}]{vosoughi2018spread}
Vosoughi S, Roy D, Aral S (2018) {The} {Spread} {Of} {True} {And} {False} {News} {Online}. \emph{{Science}} 359(6380):1146--1151.

\bibitem[{Wagner \protect\BIBand{} Wagner(1993)}]{wagner1993between}
Wagner D, Wagner F (1993) {Between} {Min} {Cut} {And} {Graph} {Bisection}. \emph{Mathematical Foundations of Computer Science 1993: 18th International Symposium, MFCS'93 Gda{\'n}sk, Poland, August 30--September 3, 1993 Proceedings 18}, 744--750 (Springer).

\bibitem[{Watchful1(2024)}]{Subreddit}
Watchful1 (2024) {Subreddit} {Comments}/{Submissions} 2005-06 {To} 2023-12. \url{https://www.reddit.com/r/pushshift/comments/1akrhg3/separate_dump_files_for_the_top_40k_subreddits/}.

\bibitem[{Yang et~al.(2022)Yang, Qureshi, \protect\BIBand{} Zaman}]{yang2022mitigating}
Yang Q, Qureshi K, Zaman T (2022) {Mitigating} {The} {Backfire} {Effect} {Using} {Pacing} {And} {Leading}. \emph{{Complex} {Networks} \& {Their} {Applications} X: {Volume} 2, {Proceedings} {Of} {The} {Tenth} {International} {Conference} {On} {Complex} {Networks} {And} {Their} {Applications} {Complex} {Networks} 2021 10}, 156--165 (Springer).

\bibitem[{Yuan et~al.(2025)Yuan, Schneider, \protect\BIBand{} Rizoiu}]{yuan_behavioral_2025}
Yuan L, Schneider PJ, Rizoiu MA (2025) Behavioral {Homophily} in {Social} {Media} via {Inverse} {Reinforcement} {Learning}: {A} {Reddit} {Case} {Study}. \emph{Proceedings of the {ACM} on {Web} {Conference} 2025}, 576--589, {WWW} '25 (New York, NY, USA: Association for Computing Machinery), ISBN 9798400712746, \urlprefix\url{http://dx.doi.org/10.1145/3696410.3714618}.

\bibitem[{Zaman et~al.(2014)Zaman, Fox, \protect\BIBand{} Bradlow}]{zaman2014bayesian}
Zaman T, Fox EB, Bradlow ET (2014) A {Bayesian} {Approach} {For} {Predicting} {The} {Popularity} {Of} {Tweets}. \emph{{The} {Annals} {Of} {Applied} {Statistics}} 8(3):1583--1611.

\bibitem[{Zarezade et~al.(2017)Zarezade, De, Upadhyay, Rabiee, \protect\BIBand{} Gomez-Rodriguez}]{zarezade_steering_2017}
Zarezade A, De A, Upadhyay U, Rabiee HR, Gomez-Rodriguez M (2017) {Steering} {Social} {Activity}: A {Stochastic} {Optimal} {Control} {Point} {Of} {View}. \emph{J. {Mach}. {Learn}. {Res}.} 18(1):7512--7546, ISSN 1532-4435.

\end{thebibliography}

\end{document}